\newif\iftwocolumn

\documentclass[final,5p,times,twocolumn]{elsarticle}
\twocolumntrue

\usepackage{graphicx}% Include figure files
\usepackage{dcolumn}% Align table columns on decimal point
\usepackage{bm}% bold math
\usepackage{xcolor}% color text
\usepackage{hyperref}
\usepackage{url}
\usepackage[mathlines]{lineno}
\usepackage[version=3]{mhchem} % Formula subscripts using \ce{}
\usepackage{amsmath}
\usepackage{amssymb}
\usepackage{amsthm}
\usepackage{changes}
\usepackage{xfrac}
\usepackage{mathtools}

\usepackage{upgreek}
\usepackage{gensymb}
\usepackage{appendix}
\usepackage{subfiles}
\usepackage{nccmath}
\usepackage{todonotes}
\usepackage{makecell}
\usepackage{diagbox}
\usepackage{textcomp}
\usepackage{float}
\usepackage{comment}
\usepackage{listings}
\usepackage{widetext}

\usepackage[T1]{fontenc}
\usepackage[utf8]{inputenc} % Ensure UTF-8 input encoding

\definecolor{matlabBlue}{rgb}{0,0,1}
\definecolor{matlabGreen}{rgb}{0,0.5,0}
\definecolor{matlabOrange}{rgb}{0.8,0.4,0}
\definecolor{matlabGray}{rgb}{0.5,0.5,0.5}

\definecolor{vscodeBlue}{RGB}{0, 0, 255}         % Keywords
\definecolor{vscodeGreen}{RGB}{0, 128, 0}        % Comments
\definecolor{vscodeRed}{RGB}{163, 21, 21}        % Strings
\definecolor{vscodeOrange}{RGB}{255, 102, 0}     % Constants, Numbers
\definecolor{vscodePurple}{RGB}{163, 73, 164}    % Functions
\definecolor{vscodeGray}{RGB}{128, 128, 128}     % Line numbers and frame
\definecolor{vscodeTextGray}{RGB}{22, 22, 22}     % Text
\definecolor{vscodeBackground}{RGB}{255, 255, 255}% Background (white)

\lstdefinestyle{MatlabEditorStyle}{
    language=Matlab,                        % Use MATLAB language definitions
    basicstyle=\ttfamily\footnotesize\color{vscodeTextGray},
    keywordstyle=\color{vscodeBlue},
    identifierstyle=\color{vscodeTextGray},
    commentstyle=\color{vscodeGreen},
    stringstyle=\color{vscodeRed},
    numberstyle=\tiny\color{vscodeGray},
    backgroundcolor=\color{vscodeBackground},
    numbers=left,
    numbersep=5pt,
    showstringspaces=false,
    frame=single,
    rulecolor=\color{vscodeGray},
    breaklines=true,
    tabsize=4,
    captionpos=b,
}

\lstdefinestyle{VSCodeStyle}{
    language=C++,
    basicstyle=\ttfamily\footnotesize\color{vscodeTextGray},
    keywordstyle=\color{vscodeBlue},
    identifierstyle=\color{vscodeTextGray},
    commentstyle=\color{vscodeGreen},
    stringstyle=\color{vscodeRed},
    numberstyle=\tiny\color{vscodeGray},
    backgroundcolor=\color{vscodeBackground},
    numbers=left,
    numbersep=5pt,
    showstringspaces=false,
    frame=single,
    rulecolor=\color{vscodeGray},
    breaklines=true,
    tabsize=4,
    captionpos=b,
    morekeywords={...},
}

\newcounter{bla}

\journal{Computer Physics Communications}

\begin{document}
%\preprint{APS/123-QED}
\begin{frontmatter}

\title{PHOENIX -- Paderborn highly optimized and energy efficient solver for two-dimensional nonlinear Schrödinger equations with integrated extensions}

\author[a,b]{Jan Wingenbach\corref{author}} %ORCID 0000-0003-3558-6972
 \address[a]{Department of Physics and Center for Optoelectronics and Photonics Paderborn (CeOPP), Paderborn University, Warburger Strasse 100, 33098 Paderborn, Germany}
 \address[b]{Institute for Photonic Quantum Systems (PhoQS), Paderborn University, Warburger Strasse 100, 33098 Paderborn, Germany}

\author[a,b]{David Bauch} % 0000-0001-5504-9619
 %\affiliation{Department of Physics and Center for Optoelectronics and Photonics Paderborn (CeOPP), Paderborn University, Warburger Strasse 100, 33098 Paderborn, Germany}
 %\affiliation{Institute for Photonic Quantum Systems (PhoQS), Paderborn University, Warburger Strasse 100, 33098 Paderborn, Germany}
 %\affiliation{GitHub Repository: \url{https://github.com/orgs/Schumacher-Group-UPB/repositories}} 
% \homepage{https://github.com/orgs/Schumacher-Group-UPB/repositories}
 
 \author[a]{Xuekai Ma}
 %\affiliation{Department of Physics and Center for Optoelectronics and Photonics Paderborn (CeOPP), Paderborn University, Warburger Strasse 100, 33098 Paderborn, Germany}

 \author[c]{Robert Schade} %https://orcid.org/0000-0002-6268-5397
 \address[c]{Paderborn Center for Parallel Computing (PC2), Paderborn University, Warburger Strasse 100, 33098 Paderborn, Germany}

  \author[c,d]{Christian Plessl}%https://orcid.org/0000-0001-5728-9982
 %\affiliation{Paderborn Center for Parallel Computing (PC2), Paderborn University, Warburger Strasse 100, 33098 Paderborn, Germany}
 \address[d]{Department of Computer Science, Paderborn University, Warburger Strasse 100, 33098 Paderborn, Germany}

\author[a,b,e]{Stefan Schumacher}
% \affiliation{Department of Physics and Center for Optoelectronics and Photonics Paderborn (CeOPP), Paderborn University, Warburger Strasse 100, 33098 Paderborn, Germany}
 %\affiliation{Institute for Photonic Quantum Systems (PhoQS), Paderborn University, Warburger Strasse 100, 33098 Paderborn, Germany}
 \address[e]{Wyant College of Optical Sciences, University of Arizona, Tucson, AZ 85721, USA}

\cortext[author] {Corresponding author.\\\textit{E-mail address:} jawi1@campus.uni-paderborn.de}

%https://www.sciencedirect.com/journal/computer-physics-communications/publish/guide-for-authors: You are required to provide a concise and factual abstract which does not exceed 250 words.
\begin{abstract}
In this work, we introduce PHOENIX, a highly optimized explicit open-source solver for two-dimensional nonlinear Schrödinger equations with extensions. The nonlinear Schrödinger equation and its extensions (Gross-Pitaevskii equation) are widely studied to model and analyze complex phenomena in fields such as optics, condensed matter physics, fluid dynamics, and plasma physics. It serves as a powerful tool for understanding nonlinear wave dynamics, soliton formation, and the interplay between nonlinearity, dispersion, and diffraction. By extending the nonlinear Schrödinger equation, various physical effects such as non-Hermiticity, spin-orbit interaction, and quantum optical aspects can be incorporated. 
PHOENIX is designed to accommodate a wide range of applications by a straightforward extendability without the need for user knowledge of computing architectures or performance optimization.
The high performance and power efficiency of PHOENIX are demonstrated on a wide range of entry-class to high-end consumer and high-performance computing GPUs and CPUs. Compared to a more conventional MATLAB implementation, a speedup of up to three orders of magnitude and energy savings of up to $99.8 \%$ are achieved. The performance is compared to a performance model showing that PHOENIX performs close to the relevant performance bounds in many situations.
The possibilities of PHOENIX are demonstrated with a range of practical examples from the realm of nonlinear (quantum) photonics in planar microresonators with active media including exciton-polariton condensates. Examples range from solutions on very large grids, the use of local optimization algorithms, to Monte Carlo ensemble evolutions with quantum noise enabling the tomography of the system's quantum state. \\

\noindent \textbf{PROGRAM SUMMARY}

\begin{small}

\noindent
{\em Program Title:} PHOENIX                                          \\
%{\em CPC Library link to program files:} (to be added by Technical Editor) \\
{\em Developer's repository link:} \url{https://github.com/Schumacher-Group-UPB/PHOENIX/} \\
{\em Licensing provision:} MIT \\
{\em Programming language:}  C++, CUDA                                 \\
%{\em Supplementary material:}                                 \\
  % Fill in if necessary, otherwise leave out.
%{\em Journal reference of previous version:}*                  \\
  %Only required for a New Version summary, otherwise leave out.
%{\em Does the new version supersede the previous version?:}*   \\
  %Only required for a New Version summary, otherwise leave out.
%{\em Reasons for the new version:*}\\
  %Only required for a New Version summary, otherwise leave out.
%{\em Summary of revisions:}*\\
  %Only required for a New Version summary, otherwise leave out.
{\em Nature of problem: Time evolution of two-dimensional nonlinear systems such as Bose-Einstein condensates, nonlinear optical systems or hybrid light-matter systems (e.g., exciton-polariton condensates).}\\%(approx. 50-250 words)
{\em Solution method: Solving the extended two-dimensional nonlinear Schrödinger equation (Gross-Pitaevskii equation) on uniformly discretized grids in real-space with a fourth-order Runge-Kutta scheme. The use of CPU and GPU and the precision used (fp32/fp64) can be set when compiling the code. Sub-grid decomposition is possible for optimal cache efficiency. The solver provides a framework that allows users unfamiliar with the details of GPU and CPU parallelization to extend the set of equations with additional terms.}\\%(approx. 50-250 words)

%{\em Additional comments including restrictions and unusual features (approx. 50-250 words):}\\
  %Provide any additional comments here.
%\begin{thebibliography}{0}
%\bibitem{1}Reference 1         % This list should only contain those items referenced in the        
%\bibitem{2}Reference 2         % Program Summary section.   
%\bibitem{3}Reference 3         % Type references in text as [1], [2], etc.
%                               % This list is different from the bibliography at the end of 
%                               % the Long Write-Up.
%\end{thebibliography}
%* Items marked with an asterisk are only required for new versions
%of programs previously published in the CPC Program Library.\\
\end{small}

\end{abstract}

%\maketitle
\end{frontmatter}

\section{Introduction}
The nonlinear Schrödinger equation describes the evolution of complex scalar fields and has applications in many areas in biology~\cite{strogatz2018nonlinear}, engineering~\cite{bongard2007automated}, mathematics~\cite{thompson2002nonlinear}, and physics~\cite{lam2003introduction}. In physics, it is commonly used to model the behavior of wave packets in nonlinear media, Bose-Einstein condensates~\cite{PhysRevLett.84.2294}, superfluids~\cite{berloff1999nonlocal}, and to investigate phenomena like solitons~\cite{malomed1996soliton}, and rogue waves~\cite{akhmediev2009rogue}. Furthermore, loss terms can be included to model dissipative systems in the field of nonlinear optics~\cite{wang2021optical} or hybrid light-matter systems such as exciton-polariton condensates~\cite{deng2010exciton, byrnes2014exciton}. In the latter case, the general dissipative nonlinear Schrödinger equation (dNLSE) reads
\begin{align}
i\hbar\partial_t\psi=\left(-\frac{\hbar^2}{2m}\nabla_{\bot}^2+V+g|\psi|^2-\frac{i\hbar\gamma}{2}\right)\psi.
    \label{eq:1}
\end{align}
Here $\psi = \psi(\textbf{r},t)$ is the complex valued scalar field in 2D in the $x-y$ plane with $|\textbf{r}|=\sqrt{x^2+y^2}$ and $\nabla_{\bot}^2=\partial_x^2+\partial_y^2$. $m$ denotes the effective particle mass. The field may be subjected to an external potential $V = V(\textbf{r})$, and $g$ determines the strength of a Kerr-type nonlinearity. The loss $\gamma$ specifies the decay of the field. Note that Eq.~\eqref{eq:1} is generally used for the study of the time evolution of the field $\psi$ in specific 2D systems. However, it could also be switched to 3D spatial systems, such as in nonlinear optics, to study the propagation of the field $\psi$ along the perpendicular $z$ direction by simply replacing the left-hand term by $i\hbar\partial_z\psi$. Since solving the nonlinear Schrödinger equation in multidimensional systems is very time-consuming, especially when additional terms are taken into account, significant attention has been paid to optimizing this process. In the recent past, solving differential equations such as variants of the Schrödinger equation on graphics processing units (GPUs) has become increasingly popular.

GPUs are devices composed of a co-processor and tightly coupled high-speed memory that have originally been designed for image processing. While the central processing unit (CPU) is optimized for data caching and flow control, GPUs are designed for highly parallel processes executed on a large number of cores. To achieve that, a large number of transistors on the GPU is devoted to data processing~\cite{fermiarchitecture}. CUDA is a parallel computing platform introduced by NVIDIA Corporation and allows GPUs to execute programs written in commonly used programming languages including C, C++, and packages like MATLAB. CUDA allows users to implement highly parallelized solvers to be used on the GPU, solving numerical problems such as large systems of coupled differential equations \cite{cudaprogrammingguide}. Thanks to modern GPU architecture, matrix operations can be performed in a fraction of the time that would be required on a CPU. In addition, the higher memory bandwidth of GPUs enables the rapid processing of large amounts of data, as it often occurs in multidimensional systems. In recent years, the use of GPUs to solve variants of the time-dependent Schrödinger equation has attracted more and more attention. This includes solving the 1D time-dependent Schrödinger equation~\cite{10.1007/978-3-031-49432-1_8}, the 2D nonlinear Schrödinger equation with a spin degree of freedom~\cite{Smith2022}, the 2D and 3D Gross-Pitaevskii equations on single and multiple GPUs~\cite{LONCAR2016406, KIVIOJA2022108427, Jiang2023, fioroni2024python}, and rotating Gross-Pitaevskii equation \cite{GAIDAMOUR2021108007}. 

\begin{figure*}[t]
    \centering
    \includegraphics[width=1.0\textwidth]{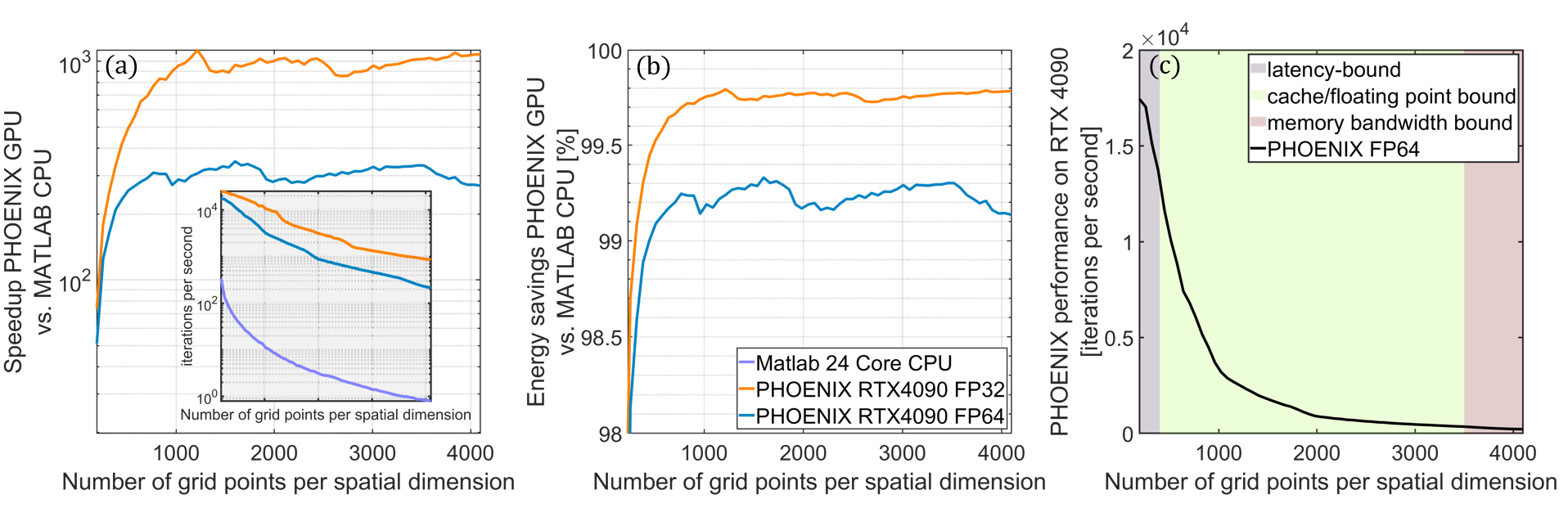}
    \caption{Performance comparison on high-end consumer hardware. Results illustrate (a) speedup and iterations per second and (b) energy savings of PHOENIX compared to a conventional Runge-Kutta CPU implementation in MATLAB for various grid sizes of a square grid, using an AMD EPYC 7443P 24core CPU and an NVIDIA RTX 4090 GPU. Simulations employ MATLAB R2024a and PHOENIX compiled with GCC and CUDA. Grid sizes ranged from $192 \times 192$ to $4096 \times 4096$, demonstrating PHOENIX's efficiency in evaluating a wide range of problems, achieving microsecond iteration times even for large grids. Performance optimization involved dividing grids into subgrids of up to $500 \times 500$ cells to minimize overhead. Both single- and double-precision calculations were performed. Energy savings are determined by the speedup and the peak power consumption for large grids manually measured using HWiNFO. (c) PHOENIX performance on RTX 4090 in double precision float and computational bounds for achievable performance according to the Roofline model introduced in Sec.~\ref{sec:perfmodel}. On the RTX 4090, PHOENIX is floating point (cache bandwidth) bound in fp64 (fp32, not shown) for intermediate grid sizes.}
    \label{fig:overview}
\end{figure*}

In these works, speedups of one~\cite{PhysRevE.91.062901, LONCAR2016406} to two~\cite{PhysRevE.91.062901, 6031567} orders of magnitude were reported. However, the works in this area restrict their equations to a few additional terms, which limits the flexibility of the code. To our knowledge, the impact of single-precision calculations on the speedup factor for the NLSE was only investigated in Ref.~\cite{fioroni2024python} to some extent. The energy efficiency of a GPU-assisted solution is rarely investigated. In the present work both CPU and GPU code are optimized to a high level to take advantage of the hardware specifics of the different architectures and to allow for a fair and transparent comparison. This opens up choices for users based on available hardware.

In the present work, we introduce the open-source code "Paderborn highly optimized and energy efficient solver for two-dimensional nonlinear Schrödinger equations with integrated extensions" (PHOENIX, \cite{phoenix_release}). This tool allows GPU-assisted solution of an extended dNLSE and is implemented in C++. We note that PHOENIX can also be executed via MATLAB or Phyton, which allows the user to process the results by familiar methods, example codes are provided~\cite{phoenix_release}. It is therefore feasible to embed PHOENIX into pre-existing code where it replaces the execution of the conventional Runge-Kutta method. PHOENIX is tailored for fast and efficient single-node CPU and single-GPU simulations because for the system sizes of interest here distributed memory is not required. However, the future extension of PHOENIX to a multi-node/multi-GPU solver is straightforward by using a space decomposition approach.
We provide a wide range of benchmarks by comparing the iteration time using the entire CPU and a single GPU. This gives a measure of resource efficiency, as well as of the speedup factor for realistic simulations. We benchmark the performance of PHOENIX based on Eq.~(\ref{eq:1}) on a variety of CPUs and GPUs from entry-level consumer to high-performance computing hardware.
%We do not use distributed memory, as this is not relevant for the system sizes of interest here. 
To illustrate the potential benefits of switching to PHOENIX for users of programming tools like MATLAB, we also compare PHOENIX to a straightforward MATLAB implementation of the fourth-order Runge Kutta method [details about the implementation are given in Sec. \ref{sec:matlab}]. The resulting time and energy savings are shown in Fig.~\ref{fig:overview} for a large range of sizes for the two-dimensional spatial grid. We run the PHOENIX code both in double and single floating-point precision. In Fig.~\ref{fig:overview} we note a speedup of up to three orders of magnitude and substantial energy savings of up to 99.8$\%$. Moreover, the comparison with the memory and cache bandwidth limits show that PHOENIX executed on the RTX 4090 is cache bandwidth bound, highlighting the cache-efficient implementation of PHOENIX. It should be noted that PHOENIX on the GPU still shows a significant speedup of one order of magnitude compared to the featured Matlab code when executed on the GPU. In addition to the rigorous performance benchmarks in the first part of the present paper, in the second part we provide explicit applications of various extensions to the implemented dNLSE as detailed below in Eqs.~(\ref{eq:PULSE_psi}) and (\ref{eq:PULSE_res}) to demonstrate the advantages and performance of PHOENIX in practical examples. 

The remainder of the present work is structured as follows: in Sec. \ref{sec:timeev} the fourth-order Runge-Kutta integration method is introduced to solve the differential Eq.~(\ref{eq:1}) on a discretized real-space grid. In Sec.~\ref{sec:matlab} the more conventional  numerical implementation in MATLAB is described. In Sec.~\ref{sec:phoenix} the numerical implementation of PHOENIX is covered. Sec.~IV contains detailed performance and efficiency results on a variety of hardware components. Sec.~\ref{sec:application} gives an overview of possible applications of the extended equations covered by PHOENIX; it also illustrates the wide range of possibilities and exceptional performance of our code with overall only moderate hardware requirements. The range of applications discussed in the present work are from the area of (quantum) photonics in planar microresonators with active media including exciton-polariton condensates. For the more use-case oriented reader, in a first read we would recommend to focus on the introductory parts of the present paper and Sec.~\ref{sec:application}. Detailed understanding of the more technical sections in between are not required for efficient use or even extensions to our code. All codes for the described applications and detailed instructions on how to use PHOENIX, including calls of the core numerical routine and pre- and post-processing of data with easy to use and customize MATLAB and Python scripts, respectively, can also be found on our Github repository~\cite{phoenix_release}.

\section{Time Evolution Schemes} \label{sec:timeev}
The time evolution of the wavefunction $\psi(\textbf{r},t)$ is governed by Eq.~(\ref{eq:1}) and following discretization on a uniform real-space grid is solved by explicit integration schemes like the Runge-Kutta integration method. We outline the process with the example of the fourth-order Runge-Kutta scheme (RK4): Let $f(t) = -\frac{\mathrm{i}}{\hbar} \partial_\mathrm{t} \psi(\textbf{r},t)$ represent the time derivative of the wavefunction for a single grid cell, and $\psi_t = \psi(\textbf{r},t)$ denote the current wavefunction value at that grid cell. The RK4 method advances the system by the time step $\Delta t$ by solving the following equations:
\begin{align}
K_1 &= f(\psi_t) \label{eq:rk4_K1} \\
K_2 &= f(\psi_t + 0.5 \Delta t K_1)  \label{eq:rk4_K2} \\
K_3 &= f(\psi_t + 0.5 \Delta t K_2)  \label{eq:rk4_K3} \\
K_4 &= f(\psi_t + \Delta t K_3)  \label{eq:rk4_K4} \\
\psi_{t + \Delta t} &= \psi_t + \frac{1}{6} (K_1 + 2K_2 + 2K_3 + K_4) \Delta t  \label{eq:rk4_final}.
\end{align}
While the standard RK4 algorithm is computationally efficient, higher-order methods like the RK-Fehlberg or Dormand-Prince can be used for systems requiring adaptive time-stepping.

We discretize the two-dimensional system into an $N_x \times N_y$ grid, where each cell corresponds to a point in the spatial domain. The time evolution of the system is computed iteratively by solving the extended dissipative nonlinear Schrödinger equation in each cell. 

\section{Numerical Implementation} \label{sec:implementation}
\subsection{MATLAB Implementation} \label{sec:matlab}
As a comparison case of how this type of simulation is conventionally handled numerically, we start by describing a typical implementation in MATLAB. A well-established method to solve partial differential equations like Eq.~(\ref{eq:1}) is the discretization of the spatial domain and approximation of the Laplacian operator via the central difference scheme. This transforms the problem into a set of coupled ordinary differential equations that can be solved explicitly in the time domain by the Runge-Kutta integration scheme. In the MATLAB implementation, we construct the Laplacian operator as a large tridiagonal sparse matrix holding nearest neighbor couplings between grid points. The integration of the differential equation then requires a series of matrix-vector multiplications. To increase computational efficiency, MATLAB's GPU computational toolbox can be used to move variables to the GPU memory. %The iteration time of the resulting code is measured by the \textit{timeit} and \textit{gputimeit} functions.

\subsection{PHOENIX}\label{sec:phoenix}
\subsubsection{Design Principles and Usage} \label{sec:design}
The PHOENIX solver presented in this work is implemented in C++ and relies on OpenMP for CPU-based parallelization and CUDA for GPU-based acceleration. Both CPU and GPU implementations share the same core code, with architecture-specific optimizations accounting for differences in memory hierarchy and execution models. PHOENIX can be easily connected to pre- and post-processing with other tools, e.g., the generation of pulse shapes or potential inputs with Python or plotting programs. Examples are included with the code~\cite{phoenix_release}.

A specific simulation is configured in PHOENIX by different means: The usage of CPU or GPU and the used precision (fp32/fp64) are set when compiling PHOENIX. Grid sizes, boundary conditions, prefactors of terms in the Hamiltonian, integrator settings, etc. are set via command line parameters. Input states, pulse shapes, and potentials can either be read from input files or generated in PHOENIX at the start of the simulation.% Examples in Sec.~\ref{sec:code_extensions} demonstrate the simplicity of extensions to our implementation.}

A primary design goal of the PHOENIX solver is to provide a framework where additional time evolution approaches or terms in the dNLSE Hamiltonian can be implemented by users unfamiliar with details of GPU and CPU parallelization approaches, while still using the hardware to its full extent. Examples in Sec.~\ref{sec:code_extensions} demonstrate the simplicity of extensions to our implementation.

\begin{figure}
    \centering
    \includegraphics[width=1.0\columnwidth]{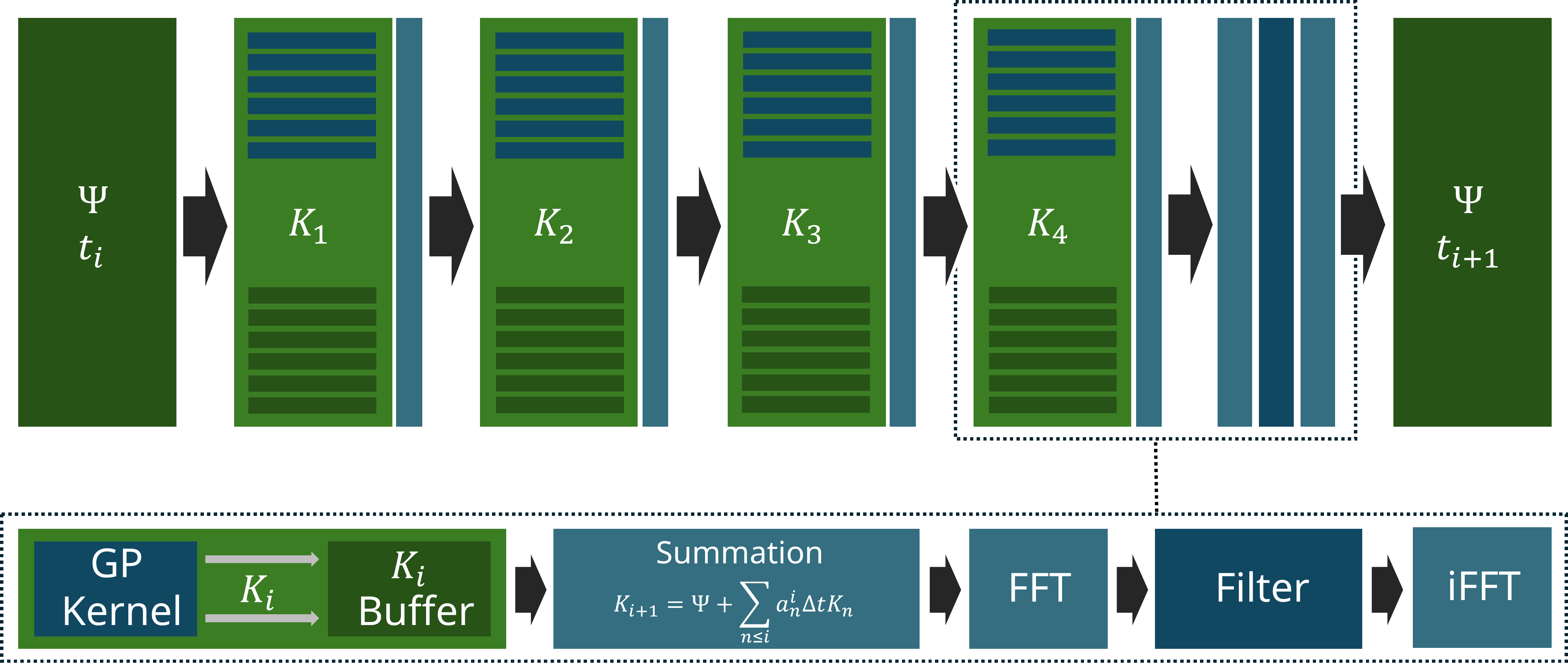}
    \caption{Schematic of the RK4 steps. Dark green objects mark real or complex buffers while blue objects mark the execution of a kernel.}
    \label{fig:gpu_implementation}
\end{figure}

In PHOENIX, the time evolution of the dNLSE is decomposed into elementary kernels so that each kernel can process all cells in parallel either with OpenMP for CPU-based parallelization or with CUDA for GPU-based acceleration. The structure of the kernel steps in Eq.~\eqref{eq:rk4_K1}-\eqref{eq:rk4_final} is schematically shown in Fig.~\ref{fig:gpu_implementation} for the example of the RK4 time step.

\subsubsection{Code Extensions}\label{sec:code_extensions}
%compare code line MATLAB to code in PHOENIX, Jan und David
%einfache erweiterung von code extensions erwähnen, eventuell Beispiel
% show example RK4 in code

PHOENIX is designed for straightforward code extensions. Code Listing~\ref{lst:extintegrator} presents the implementation of the RK4 time-stepping algorithm in PHOENIX, corresponding to Eq.~\eqref{eq:rk4_K1}–\eqref{eq:rk4_final}. Due to its expressive and compact formulation, other integrators can be implemented without requiring background knowledge of GPU/CPU parallelization or prior experience with C++ programming.

In addition to alternative integrators, simulations may necessitate the inclusion of different terms in the Hamiltonian defined by Eq.~\ref{eq:1}. Such terms can be seamlessly added to the kernel, as illustrated in Code Listing~\ref{lst:extterm}.

The extensibility of time-stepping algorithms and Hamiltonian terms in PHOENIX is as straightforward as in conventional MATLAB code.

%As we compare our custom implementation with a more conventional MATLAB code, here we also provide a MATLAB code in Listing~\ref{lst:extintegrator_matlab}. It is noteworthy that adding additional terms to the equations in PHOENIX is as straightforward in code as it is in MATLAB. Similarly, extending the RK4 integrator to higher or lower orders is intuitively achieved using predefined macros, as demonstrated in Listing~\ref{lst:extintegrator}.

%\begin{widetext}

\begin{lstlisting}[language=C++,style=VSCodeStyle,caption={Sample implementation of the RK4 time step in PHOENIX. This highlights how easily modifications can be made to the existing code.},
  morekeywords={int,char,double,float,unsigned,CALCULATE_K,INTERMEDIATE_SUM_K,FINAL_SUM_K,SOLVER_SEQUENCE},
  label={lst:extintegrator}
]
// Calculate K1 from input "wavefunction"
CALCULATE_K( 1, wavefunction );
// Calculate Psi + dt/2*K1
// Saved into "buffer_wavefunction"
INTERMEDIATE_SUM_K( 1, 0.5f );
// Calculate K2 from input "buffer_wavefunction"
CALCULATE_K( 2, buffer_wavefunction );
// Calculate Psi + dt/2*K2
// Saved into "buffer_wavefunction"
INTERMEDIATE_SUM_K( 2, 0.5f );
// Calculate K3 from input "buffer_wavefunction"
CALCULATE_K( 3, buffer_wavefunction );
// Calculate Psi + dt*K3
// Saved into "buffer_wavefunction"
INTERMEDIATE_SUM_K( 3, 1.0f );
// Calculate K4 from input "buffer_wavefunction"
CALCULATE_K( 4, buffer_wavefunction );
// Sum all K's into new Psi
FINAL_SUM_K( 4, 1.0f/6.0f, 1.0f/3.0f,
                1.0f/3.0f, 1.0f/6.0f );
\end{lstlisting}

\begin{lstlisting}[language=C++,style=VSCodeStyle,caption={Implementation of the terms $g |\Psi|^2$ and $-i\hbar\gamma/2$ in PHOENIX.},
  directivestyle={\color{black}},
  morekeywords={abs2,mi_conjugate,real,complex},
  emphstyle={\color{blue}},
  label={lst:extterm}
]
const Type::real in_psi_norm = CUDA::abs2( in_wf ); //|Psi|^2
const Type::complex in_wf_mi = CUDA::mi_conjugate( in_wf ); //-i*in_wf
wf_plus += one_over_h_bar_s * g_c * in_psi_norm * in_wf_mi;
wf_plus -= Type::real( 0.5 ) * gamma_c * in_wf;
\end{lstlisting}

\begin{comment}
\begin{lstlisting}[language=Matlab,style=MatlabEditorStyle,caption={Implementation of RK4 time step as well as terms $g |\Psi|^2$ and $-i\hbar\gamma/2$ in Matlab for comparison.},
  label={lst:extintegrator_matlab}
]
...
% |Psi|^2
abs2y = abs(y).^2;
% Individual K is calculated using Eq. 1
k1 = -1i/hbar*(H*y1 + gc*abs2y.*y1 + 1i*hbar/2*(-gammaC*y1));
% Calculate intermediate arguments 'temp' and new |temp|^2
temp = y1+dt*k1/2; abs2y = abs(temp).^2;
% Repeat until K4
k2 = -1i/hbar*(H*temp + gc*abs2y.*temp + 1i*hbar/2*(-gammaC*temp));
temp = y1+dt*k1/2; abs2y = abs(temp).^2;
k3 = -1i/hbar*(H*temp + gc*abs2y.*temp + 1i*hbar/2*(-gammaC*temp));
temp = y1+dt*k1; abs2y = abs(temp).^2;
k4 = -1i/hbar*(H*temp + gc*abs2y.*temp + 1i*hbar/2*(-gammaC*temp));
% Sum all K's into new Psi
y = y + dt*(k1 + 2*k2 + 2*k3 + k4)/6;
...
\end{lstlisting}
\end{comment}
%\end{widetext}

\subsubsection{Smoothing} \label{sec:smoothing}
Finally, a Fast Fourier Transform (FFT) can be applied to $\psi$ to filter the resulting wavefunction in $k$-space, followed by an inverse FFT to return the data to real space. We use the library cuFFT~\cite{cudaprogrammingguide} in GPU and FFTW~\cite{FFTW05} in CPU calculations. For systems with spin degrees of freedom, equivalent steps are applied for the two spinor components $\psi_\pm$. This filtering may be necessary to suppress numerical artifacts forming due to various reasons, including certain shapes of external pumping sources and boundary effects in $k$-space.

\subsubsection{Subgrid Decomposition} \label{sec:subgrid}
The computations required for the time steps, i.e., the evaluation of the derivative $f(t) = -\frac{\mathrm{i}}{\hbar} \partial_t \psi(\textbf{r},t)$ and the scaling/additions in the five steps of the RK4 schema in Eq.~\eqref{eq:rk4_K1}-\eqref{eq:rk4_final} have a low computational intensity like most other numerical stencil problems. A low computational intensity means that for the required floating-point operations a comparatively large amount of data needs to be transferred from the caches or memory. Thus, typically the bottleneck of the calculation is not the floating-point throughput but the useable bandwidth of caches or the bandwidth of the memory interface. The bandwidth of the memory interface of GPUs or CPUs is often up to an order of magnitude lower than the usable cache bandwidths. Hence, optimizing a code to use the caches more efficiently can deliver a drastic speed-up in situations with low computational intensity like the time evolution of the dNLSE in this work.

For this purpose, we have implemented a subgrid decomposition where the primary $N_x \times N_y$ grid is decomposed into smaller subgrids of size $N_{x,\mathrm{sg}} \times N_{y,\mathrm{sg}}$. Each subgrid includes a \textit{halo} of grid cells corresponding to neighboring subgrids so that the calculations for the time step, i.e., Eq.~\eqref{eq:rk4_K1}-\eqref{eq:rk4_final}, is independent for each subgrid. Thus, during a time step on a subgrid instead of the full grid of $N_x \times N_y$ points only the points of a subgrid, i.e., $N_{x,\mathrm{sg}} \times N_{y,\mathrm{sg}}$ need to be processed in the individual steps making the caches much more efficient due to the reduced data size. For a more readable notation, we assume that the grid dimensions are multiples of the subgrid dimensions for each spatial dimension, i.e., $N_x/N_{x,\mathrm{sg}}=n_{x,\mathrm{sg}}\in \mathbb{N}$ and $N_y/N_{y,\mathrm{sg}}=n_{y,\mathrm{sg}}\in \mathbb{N}$. The solver can also handle situations where this is not the case.

It should be noted that the halo region around the subgrids introduces a slight overhead in terms of the number of computed cells in a time step because instead of $N_{x,\mathrm{sg}}\times N_{y,\mathrm{sg}}$ cells per subgrid now $(N_{x,\mathrm{sg}}+2N_\mathrm{halo})\times (N_{y,\mathrm{sg}}+2N_\mathrm{halo})$ cells per subgrid need to be computed. 

After a time step has been applied to each subgrid individually, a halo exchange step is introduced in which boundary values are synchronized by copying updated boundary values between neighboring subgrids. This technique also enables efficient parallel execution on CPUs using OpenMP, with each subgrid being assigned to a separate CPU core such that the number of subgrids $n_\mathrm{sg}$ is a multiple of the number of CPU cores.

\subsubsection{Lower Precision Floating-Point Formats}\label{sec:lowprec}
While the use of double-precision floating point numbers (fp64) was prevalent in computational simulations in the past, large improvements in throughput and computational efficiency can be achieved in situations where lower precision calculations are possible or the lower precision can be compensated by changes in the algorithms~\cite{10.5555/3291656.3291719,doi:10.1137/18M1229511,Higham_Mary_2022,SCHADE2022102920,doi:10.1177/10943420211003313}. A straightforward reduction from double precision to single precision (fp32) in the simulation of the dNLSE with RK schemas is accurate except (for our examples) in cases where sharp spectral resonances appear and need to be matched to a high precision [see Sec.\ref{cha:ex1}]. Thus, the solver presented in this paper can use either single or double precision for the calculation of the time evolution steps, Eqs.~\eqref{eq:rk4_K1}-\eqref{eq:rk4_final}. The step from double precision to single precision already gives a number of advantages: On the one hand, all data objects are reduced in size by a factor of two so that transfer times from caches or memory are reduced. On the other hand, consumer GPUs typically only have a low double-precision performance because single-precision calculations are sufficient for graphical applications. Thus, using single-precision instead of double-precision formats ensures that the floating point throughput doesn't bottleneck the solver on consumer-grade GPUs.

\subsubsection{Data Volume} \label{sec:data_volume}
The data required for example in the GPU memory consists of buffers of complex single- or double-precision floating-point numbers with 
$n_{x,\mathrm{sg}}\times n_{y,\mathrm{sg}} \times (N_{x,\mathrm{sg}}+2N_\mathrm{halo})\times (N_{y,\mathrm{sg}}+2N_\mathrm{halo})$ elements, where $n_{x,\mathrm{sg}}$ and $n_{y,\mathrm{sg}}$ are the numbers of subgrid dimensions in $x$ and $y$ dimension, respectively.
For scalar systems with only the wavefunction $\psi$ six buffers suffice. However, more complex systems, such as those with a reservoir (see example in Section~\ref{cha:ex2}), require an additional six buffers, and systems including spin-orbit interaction (see example in Section~\ref{cha:ex3}) require 24 buffers in total.

Thus, the required CPU main memory or GPU memory for a scalar simulation without a reservoir using RK4 ($N_\mathrm{halo}=4$) in single precision can be determined as
\begin{eqnarray}
    M=48 n_{x,\mathrm{sg}}\times n_{y,\mathrm{sg}} &\times& (N_{x,\mathrm{sg}}+2N_\mathrm{halo})\nonumber\\&\times& (N_{y,\mathrm{sg}}+2N_\mathrm{halo}) \ \mathrm{byte} \label{eq:activedata}
\end{eqnarray}
Thus except for extremely large simulations with grid sizes beyond $\approx 10^4$, the memory requirement doesn't limit the choice of, for example, consumer GPUs.

\subsubsection{Vectorization and Parallelization} \label{sec:vectorization}
GPU parallelization consists of individual kernels that loop over subgrids of cells. The kernel launch latencies are mitigated by using CUDA graphs. In the GPU implementation, no special consideration of the storage format of complex numbers is required. So whether the complex numbers are stored as interleaved real and imaginary parts or separate arrays for the real and imaginary parts doesn't need additional attention. Thus in practice, the interleaved format is preferred due to the better readability of the program code. In contrast, for CPUs the interleaved format leads to challenges with vectorization if complex conjugation or multiplications are used. In such situations, the available compilers can often not sufficiently vectorize the code which can lead to unvectorized floating-point throughput becoming a bottleneck. For PHOENIX this problem is circumvented with temporary buffers for example for $-i\psi_t$ which by itself is computed in an efficiently vectorized way for example with AVX2 compiler intrinsics.

\subsubsection{Performance Modeling} \label{sec:perfmodel}
The time step in Eq.~\eqref{eq:rk4_K1}-\eqref{eq:rk4_final} requires three types of kernels on the grid. If multiple subgrids are used, also the halo exchange kernel is required. We limit the performance modeling to the scalar case without a reservoir and no smoothing.

The derivative kernel, i.e., $f(t)=-\frac{1}{\hbar} \nabla_t \psi(\vec r,t)$ used in~\eqref{eq:rk4_K1}-\eqref{eq:rk4_K2} requires the evaluation of the right side of Eq.~\eqref{eq:1} at the grid points $\mathbf{r}_{i,j}=(i\Delta x,j\Delta y)$ for $\psi(i,j)=\psi(\mathbf{r}_{i,j},t)$ and most importantly a five-point stencil
%\begin{widetext}
\begin{align}
    \nabla^2 \psi(i,j)=\frac{1}{\Delta x^2}[\psi(i+1,j)+\psi(i-1,j)-2\psi(i,j)] \nonumber \\ +\frac{1}{\Delta y^2}[\psi(i,j+1)+\psi(i,j-1)-2\psi(i,j)]
\end{align}
%\end{widetext}
for the spatial derivative. This stencil will effectively require three reads of the wavefunction $\psi$ because $\psi(i+1,j)$, $\psi(i-1,j)$, and $\psi(i,j)$ are contained in one cache line or adjacent cache lines whereas $\psi(i,j+1)$ and $\psi(i,j-1)$ require additional reads of cache lines.

The two-term summation kernels used in Eq.~\eqref{eq:rk4_K2}-\eqref{eq:rk4_K4} with the structure
\begin{align}
    z=x+\alpha y,
\end{align}
are identical to the well-known STREAM triad~\cite{McCalpin1995,McCalpin2007} case. The five-term summation kernel in Eq.~\eqref{eq:rk4_final} with the structure
\begin{align}
    z=x+\alpha_1 y_1+\alpha_2 y_2+\alpha_3 y_3+\alpha_4 y_4,
\end{align}
can be seen as a generalization of the STREAM triad case.

While the summation kernels could be integrated into the call of the derivative kernel, we maintain a separation to allow flexibility and straightforward extendability, such as implementing higher-order Runge-Kutta methods. 

Table~\ref{tab:readswrites} lists the required effective read/write counts of buffers in the kernels and the computational intensity, i.e., the ratio of performed floating-point operations to data transfers.

The computational intensity $B$ of the derivative kernel depends on the terms on the right side of Eq.~\eqref{eq:1}. We have assumed the case of $g\neq 0$ and $\gamma \neq 0$. 

\begin{table}[h!] 
    \centering 
    \begin{tabular}{|c|c|c|c|} 
        \hline 
       Kernel & \makecell{Number of\\ Reads $n_r$} & \makecell{Number of\\ Writes $n_w$} & \makecell{Comp. Intensity\\ $B$ (Flops/byte)}  \\ \hline 
        \makecell{derivative\\ kernel} & 3 & 1 & $\approx \frac{1}{2}$ \\ \hline %4, 20 flops per cell, 2*4*4 byte per cell
        \makecell{two-term\\ sum} & 2 & 1 & $\frac{1}{6}$ \\ \hline %3, 2*2 flops per cell, 2*4*3 byte per cell
        \makecell{five-term\\ sum} & 5 & 1 & $\frac{1}{4}$\\ \hline %1, 2*6 flops per cell, 2*4*6 byte per cell
        \makecell{RK4\\ iteration} & 23 & 8 & $\approx \frac{1}{3}$ \\ \hline 
    \end{tabular} 
    \caption{Effective expected numbers of reads $n_r$ and writes $n_w$ as well as the computational intensity for the kernels for the scalar case without a reservoir (cf. examples below in Section~V) in single precision.} 
    \label{tab:readswrites}
\end{table}
To put the computational intensities into perspective it is useful to consider the properties of typical hardware listed in Table~\ref{tab:hardware} that are characterized by the peak floating-point throughput $P^\mathrm{fp32/64}_\mathrm{max}$ in single or double precision and the cache/memory transfer bandwidths $b^\mathrm{cache/mem}_\mathrm{max}$.
If the computational intensity $B$ of the code is larger than the computational intensity $B_\mathrm{H}^\mathrm{fp32/64}=P^\mathrm{fp32/64}_\mathrm{max}/b^\mathrm{cache/mem}_\mathrm{max}$ of the hardware then the code is limited by the floating-point throughput, otherwise it is limited by transfer bandwidth. For all hardware listed in Table~\ref{tab:hardware} the solver is estimated to be transfer bound in single precision.
 
\begin{table}[h!] 
    \centering 
    \begin{small}
    \begin{tabular}{|c|c|c|c|c|c|} 
        \hline 
        Hardware & \makecell{Cache\\ Size\\ {\footnotesize(MiB)}} & \makecell{$b_\mathrm{max}^\mathrm{cache}$\\ {\footnotesize(GB/s)}} & \makecell{$b_\mathrm{max}^\mathrm{mem}$\\ {\footnotesize(GB/s)}} & \makecell{$P_\mathrm{max}^\mathrm{fp32}$\\ {\footnotesize(TFlop/s)}}& \makecell{$P_\mathrm{max}^\mathrm{fp64}$\\ {\footnotesize(TFlop/s)}} \\ \hline
        \makecell{NVIDIA RTX\\ 4060 TI 16 GB} & 32 & 1600 & 275 & 22  & 0.3 \\ \hline
        \makecell{NVIDIA RTX\\ 4090 24 GB} & 72 & 5300 & 950 & 82.6  & 1.3 \\ \hline
        \makecell{NVIDIA H100\\ SXM5 80 GB} & 50 & 8000 & 3200 & 66.9 & 33.5 \\ \hline
        AMD 5800X3D & 96 & 580 & 31 & 1.1 & 0.5 \\ \hline
        \makecell{AMD 7763\\ CCD, 8 cores} & 32 & 520 & 38 & 0.9  & 0.4 \\ \hline
        \makecell{AMD 7763\\ full, 64 cores} & 256 & 2600 & 155 & 2.5 & 5.1  \\ \hline
    \end{tabular} 
    \end{small}
    \caption{Hardware specifications of the used CPUs and GPUs. For CPUs, the peak cache bandwidths $b^\mathrm{cache}_\mathrm{max}$ (L3), memory bandwidths $b^\mathrm{mem}_\mathrm{max}$, and the peak floating-point throughputs $P^\mathrm{fp32/64}_\mathrm{max}$ in single (SP) and double (DP) precision have been measured LIKWID (5.3, ~\cite{7103452}, triad\_avx\_fma, triad\_mem\_avx\_fma, peakflops\_avx\_fma, peakflops\_sp\_avx\_fma). For NVIDIA GPUs, the maximal useable cache/memory bandwidths have been measured with gpu-benches (\cite{gpu-benches}, gpu-stream/gpu-l2-cache), and non-tensor floating-point throughputs are according to the device specifications~\cite{hopperwhitepaper,adawhitepaper}. The AMD 7763 CPU consists of 8 core complex dies (CCDs) with 8 CPU cores each.} 
    \label{tab:hardware}
\end{table}

Using the Roofline model the maximal performance of the program can be estimated by considering the maximal useable floating-point throughput $P_\mathrm{max}$ and maximal useable memory bandwidth $b^\mathrm{mem}_\mathrm{max}$ or cache transfer bandwidth $b^\mathrm{cache}_\mathrm{max}$ in relation to the computational intensity $B$ of the code. The assumptions of the Roofline model that either the floating-point throughput or the transfers are the bottlenecks lead to the performance estimate
\begin{align}
    P^\mathrm{mem/cache}=\min(P_\mathrm{max},B\cdot b^\mathrm{cache/mem}_\mathrm{max})\label{eq:roofline}
\end{align}
for the floating-point performance assuming that either memory or cache are the relevant transfer stages for the buffers. This estimate excludes the halo synchronization step which would require a more elaborate treatment. 

\begin{figure}
    \centering
    \includegraphics[width=1.0\linewidth]{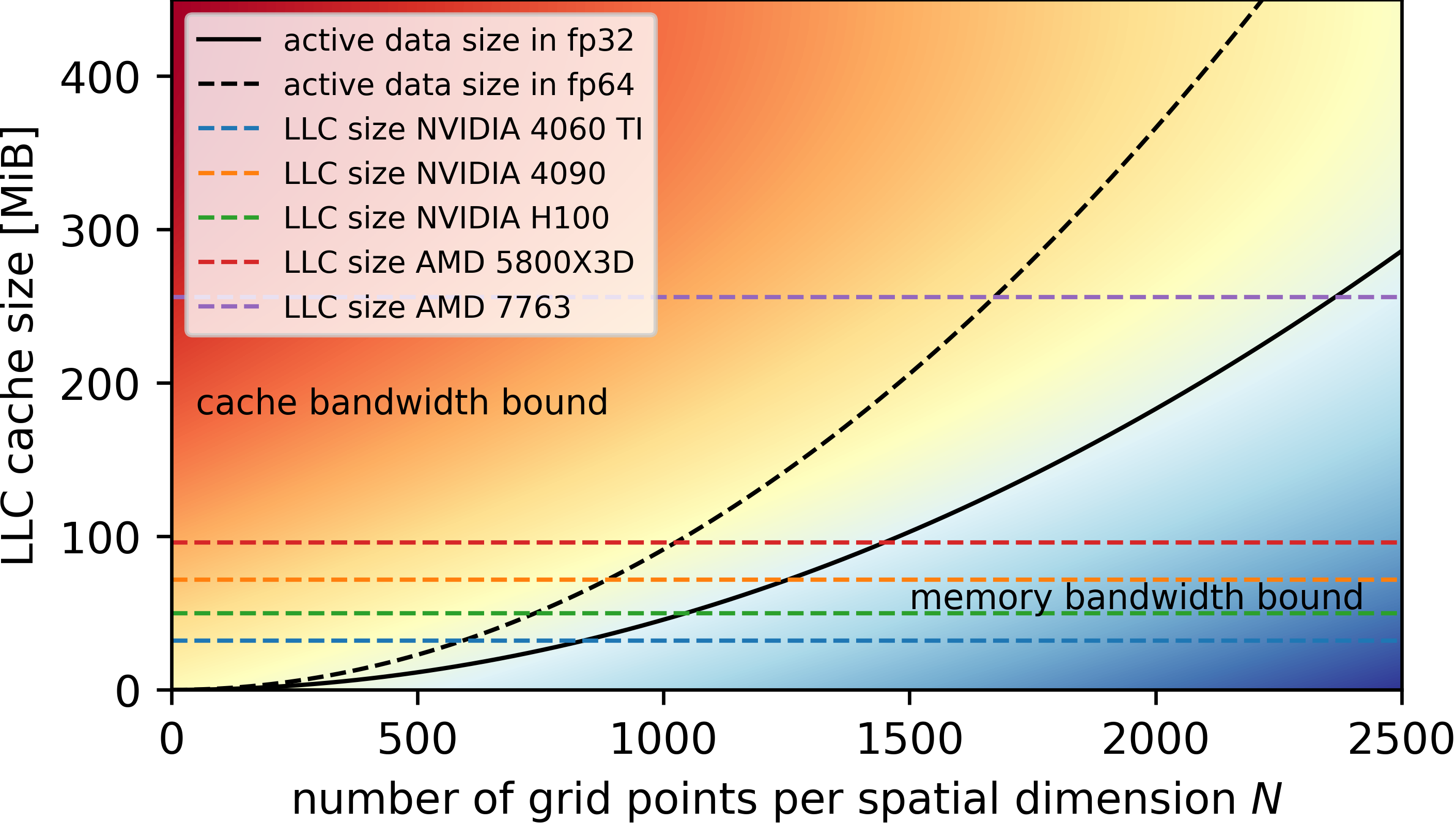}
    \caption{The last-level cache (LLC) sizes of the devices in Tab.~\ref{tab:hardware} is shown in comparison to the active data size, Eq.~\eqref{eq:activedata}, for single precision (fp32) and double precision (fp64) calculations of $N\times N$ grids. The cache-bandwidth-bound regions (red) and memory-bandwidth-bound regions (blue) are shown as shaded areas.}
    \label{fig:perfmodel_bounds}
\end{figure}

Which transfer stage, either one of the cache levels or main memory, is relevant depends on the size of data being processed in parallel. Fig.~\ref{fig:perfmodel_bounds} compares the active data size given by Eq.~\eqref{eq:activedata} to the last-level cache (LLC) sizes of the devices in Tab.~\ref{tab:hardware}. If the data size is smaller than the cache size, the cache bandwidth determines the maximal achievable performance in Eq.~\eqref{eq:roofline} otherwise the memory bandwidth is the relevant bandwidth. The subgrid decomposition described in Sec.~\ref{sec:subgrid} increases the cache locality within each time step.% but not the overall size of active data.

The Roofline model predicts for hardware with $b_\mathrm{max}^\mathrm{mem}=1$ TB/s memory bandwidth like high-end consumer GPUs that the solver performs about 330 GFlops/s and, thus, would not be floating-point-throughput bound in single precision on a typical consumer GPU but transfer bandwidth bound.
For a more immediate use and comparison, the execution time of a time step can be estimated for the scalar case without a reservoir to be at least
\begin{align}
t_\mathrm{iter,roofline}&=\frac{2\cdot (n_r+n_w)N_xN_y\cdot d}{b_\mathrm{max}}\\
&=\begin{cases}
\frac{248\ \mathrm{byte}\cdot N_xN_y}{b_\mathrm{max}}, \mathrm{single} \ \mathrm{prec.}\\
\frac{496\ \mathrm{byte}\cdot N_xN_y}{b_\mathrm{max}}, \mathrm{double} \ \mathrm{prec.}
\end{cases}
\end{align}
according to the Roofline model in the transfer-bound case, where $n_r$ and $n_w$ are the numbers of reads and writes given in Tab.~\ref{tab:readswrites} and $d$ is the data size of the floating-point number format. For other cases, for example with a reservoir, the additional buffers described in Sec.~\ref{sec:data_volume} have to be considered.

\section{Performance and Efficiency Results} \label{sec:perfres}
In this section, detailed performance results are presented to evaluate the PHOENIX implementation in relation to the performance estimates of the performance model in Sec.~\ref{sec:perfmodel}.

All benchmark runs in this chapter have been performed with GCC 13.2.0 and CUDA 12.6. Detailed documentation of the environments can be found in the benchmark documentation included in the PHOENIX release~\cite{phoenix_release}.

The main performance indicator is the number $U$ of updated points per second (UP/s). It is related to the runtime $t_{iter}$ of timestep for a given grid size $N_x \times N_y$ as 
\begin{align}
t_\mathrm{iter}=N_xN_y/U.
\end{align}
The scalar case without a reservoir and smoothing is considered here for quadratic grids with the grid size $N$, i.e., $N_x=N_y=N$ with $N_x\cdot N_y=N^2$ total grid points. Hence, the total data size per grid point in all required buffers is $48$ byte in single precision and $96$ byte in double precision. 

The grid point update rates $U$ measured on the devices listed in Tab.~\ref{tab:hardware} are shown in Fig.~\ref{fig:perf_subgrid} for different choices of subgrids and floating-point precisions. In addition, the grid point update rates estimated with the roofline model are shown for comparison.
The performance of PHOENIX on CPUs and GPUs shows three characteristic regions:
\begin{description}
    \item[Small grid sizes] The simulation is latency-bound due to parallelization overhead from kernel launches on GPUs and OpenMP-parallelization loops on CPUs.
    \item[Intermediate grid sizes] The highest performance is achieved because the simulation can efficiently utilize the cache.
    \item[Large grid sizes] The memory bandwidth of the device limits the performance.
\end{description}

\subsection{Comparison to Roofline Bounds}
The performance bounds due to the cache bandwidths and memory bandwidths are shown in Fig.~\ref{fig:perf_subgrid} in comparison to the measured performance results. GPU-based simulations with PHOENIX can reach between 51 \% (NVIDIA H100, Hopper architecture) and 94 \% (NVIDIA 4060 TI, Ada Lovelace architecture) of the cache bandwidth bound for intermediate grid sizes and saturate the memory bandwidth bound for large grid sizes in single precision. In double precision, the situation is similar for the NVIDIA H100 with 54 \% of the cache bandwidth bound but different for the NVIDIA 4060 TI and NVIDIA 4090 where the floating-point throughput bound limits the performance to $40$ \% of the cache bandwidth bound resp. 70 \% of the floating-point bound.
In all other cases, the comparison from single precision (fp32) simulations to double precision simulations (fp64) shows a difference of a factor of two for investigated GPUs and CPUs due to the doubling of the data size.

CPU-based simulations with PHOENIX reach 35-62 \% of the cache-transfer bound in cache-bandwidth bound situations and saturate the memory bandwidth for large grid sizes.

\subsection{Cache-Size Dependence}
\begin{figure}
    \centering
    \iftwocolumn
    \includegraphics[width=1.0\linewidth]{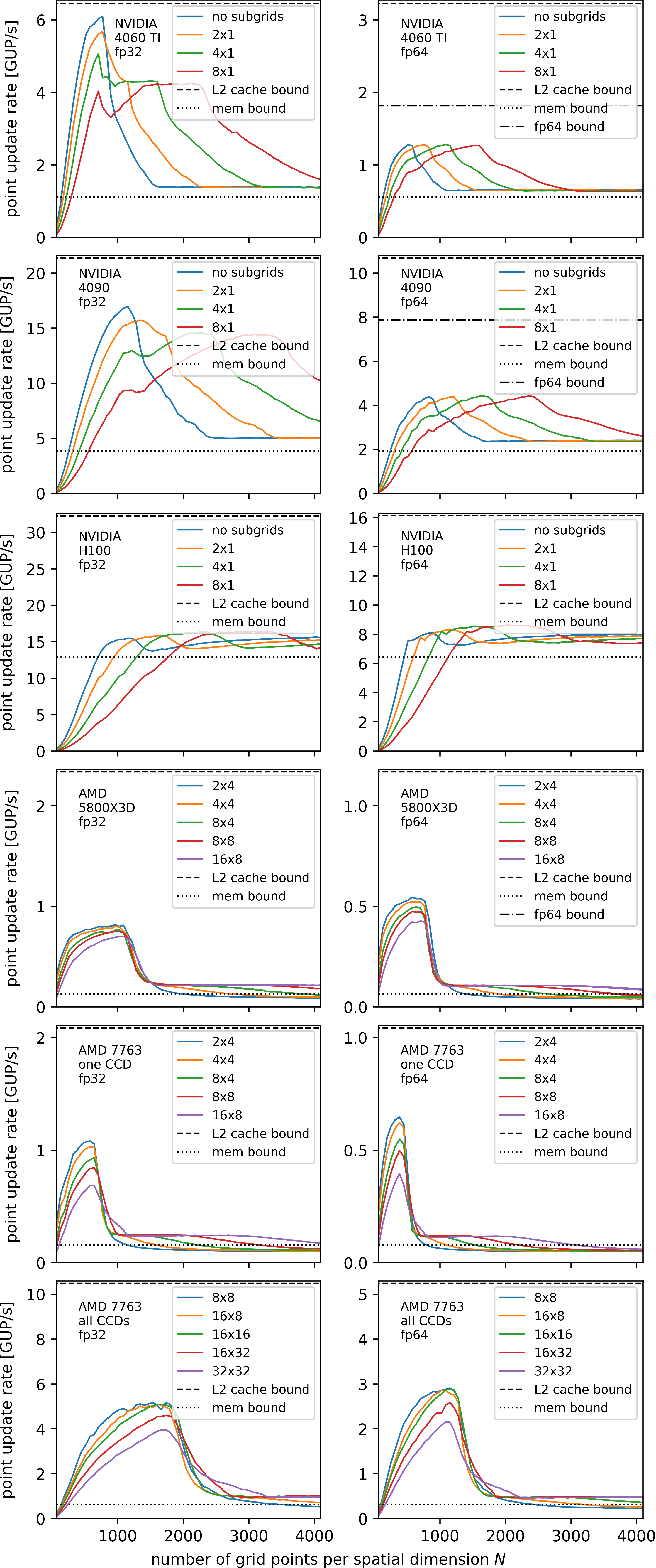}
    \else
    \includegraphics[width=0.85\linewidth]{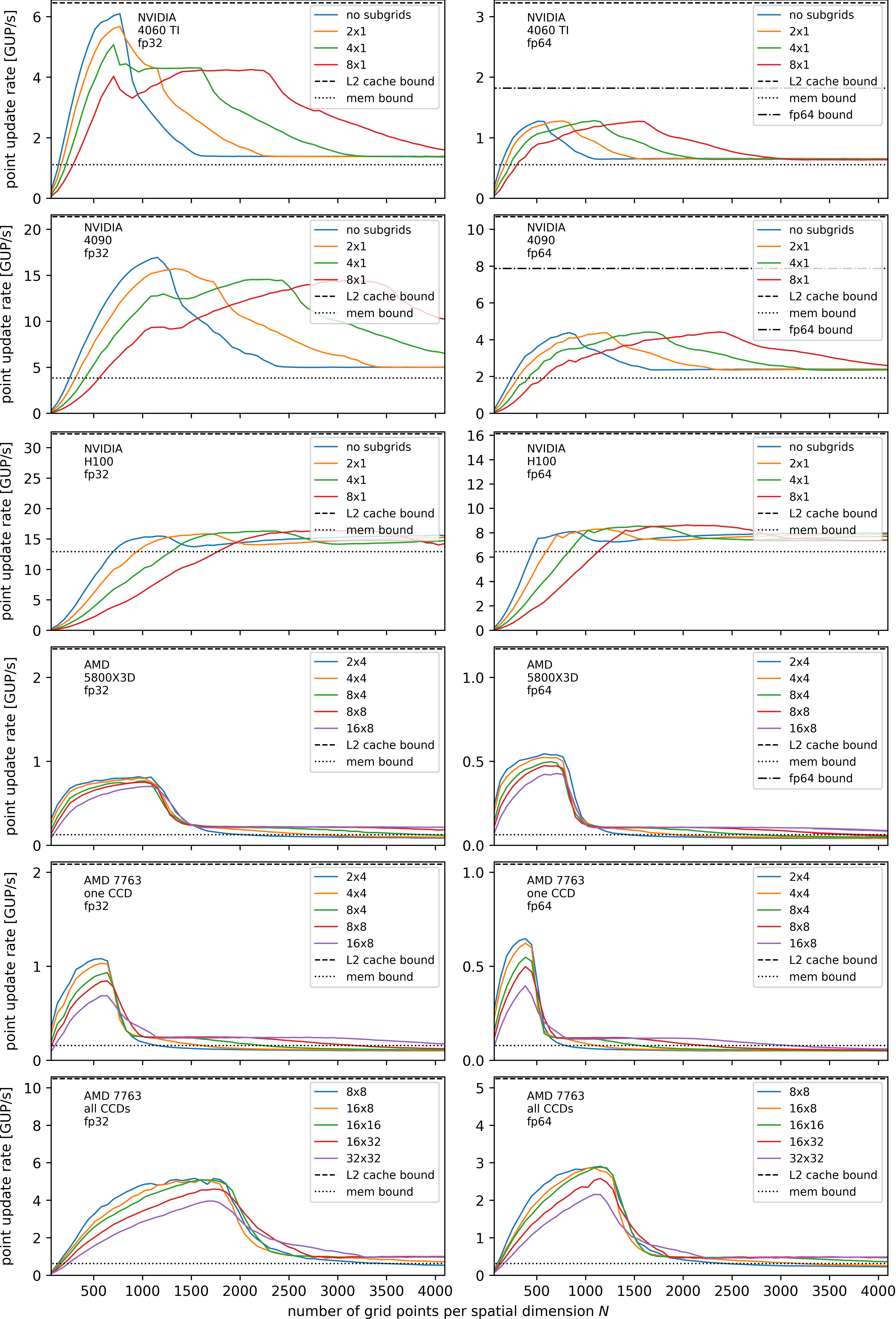}
    \fi
    \caption{The grid point update rates measured on the devices listed in Tab.~\ref{tab:hardware} are shown for simulations in single precision (fp32, left column) and double precision (fp64, right column) for different subgrid settings of the $N\times N$ grid. The corresponding performance estimate according to the roofline model for the cache bandwidth bound (dashed lines) and memory bandwidth bound (dotted lines) cases are shown for comparison.
    }
    \label{fig:perf_subgrid}
\end{figure}

The extent of regions with high performances at intermediate grid sizes, e.g., $500\leq 1500$ for the NVIDIA 4060 TI (fp32, no subgrids) directly depends on the size of the last-level cache of the devices. This effect is especially pronounced for the consumer-grade GPUs and the CPUs because of the high ratio between cache bandwidths and memory bandwidths in these devices.
A comparison between the case of using just one CCD of the AMD 7763, i.e., 8 CPU cores to the AMD 5800X3D (8 CPU cores) shown in the last two rows of Fig.~\ref{fig:perf_subgrid} demonstrates the cache size-effect because the number and type of used CPU cores (AMD Zen3) as well as the effective memory bandwidth of the cores is very similar. The region of high-performing intermediate grid sizes is extended from $N\approx 700$ for the AMD 7763-CCD (32 MB L3 cache) to $N\approx 1200$ for the AMD 5800X3D (96 MB L3 cache).

\subsection{Subgrid Dependence}
Decomposing the grid into more subgrids leads to similar effects on GPUs and CPUs by reducing the performance of small grid sizes but increasing the performance of large grid sizes. 
For small grid sizes, a higher number of subgrids increases the parallelization overhead. The computational overhead added by the subgrid halos increases this effect because the ratio of halo points to subgrid points is unfavorable for small sizes of subgrids.

For large grid sizes, the higher cache locality from the subgrid decomposition increases the performance so that the memory bandwidth bound limits only for larger grid sizes. This effect is especially noticeable for the NVIDIA 4060 TI and NVIDIA 4090 where in single precision the sub-grid decomposition can lead to a roughly threefold increase in performance.

In contrast to GPUs, the CPUs show an overall weaker dependence on the number of subgrids. One reason is that at least one subgrid per CPU core is required for the simulation in the first place and the number of subgrids is required to be a multiple of the number of CPU cores for a balanced computational load. 

\subsection{Kernel Performance} \label{sec:perf_kernel}
\begin{figure}
    \centering
    \includegraphics[width=1.0\linewidth]{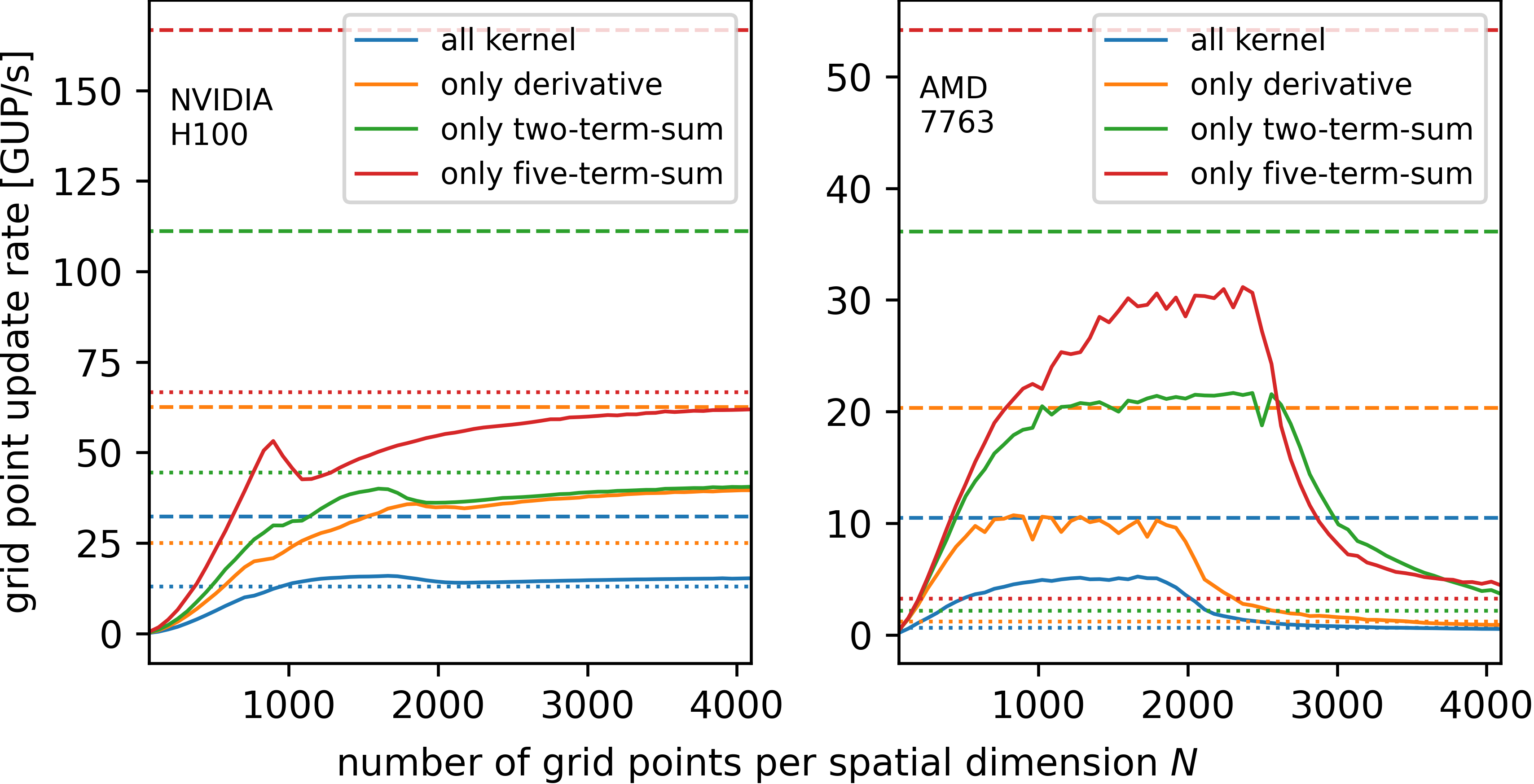}
    \caption{The grid point update rates for single-precision RK4-time steps where only the individual kernels given in Tab.~\ref{tab:readswrites} were active for the NVIDIA H100 GPU (2x1 subgrids) and the AMD 77632 CPU (8x8 subgrids). Dashed lines show the corresponding cache-bandwidth-bound performance estimates and dotted lines show the corresponding memory-bandwidth-bound performance estimates.}
    \label{fig:perf_kernel}
\end{figure}

In this section, the kernels are considered individually to identify the reason for the difference between the roofline model estimates in Sec.~\ref{sec:perfmodel} and the measured performance in Fig.~\ref{fig:perf_subgrid}. Fig.~\ref{fig:perf_kernel} compares the update rates of RK4-time steps that had all kernels except one deactivated to the corresponding estimates from the corresponding cache bandwidth bounds and memory bandwidth bounds in single precision. 

For large grid sizes, most of the individual kernel performances approach the estimates from the memory bandwidth bounds. One exception is the derivative kernel on the H100 which shows a higher than expected performance. For intermediate grid sizes where the discrepancy between the cache-bandwidth bound and measured performance was observed in Fig.~\ref{fig:perf_subgrid}, on the NVIDIA H100 the two- and five-term-sum kernels show a large performance difference compared to the cache bandwidth bound. This is in stark contrast to the NVIDIA 4060 TI and NVIDIA 4090 GPUs where this behavior is not observed. The detailed investigation and optimization of the cache efficiency of PHOENIX is a topic for future research. For the CPU execution, all three kernels perform similarly when compared to the corresponding cache bandwidth bound.

\subsection{Energy Usage} \label{sec:perf_energy}
\begin{figure}
    \centering
    \includegraphics[width=1.0\linewidth]{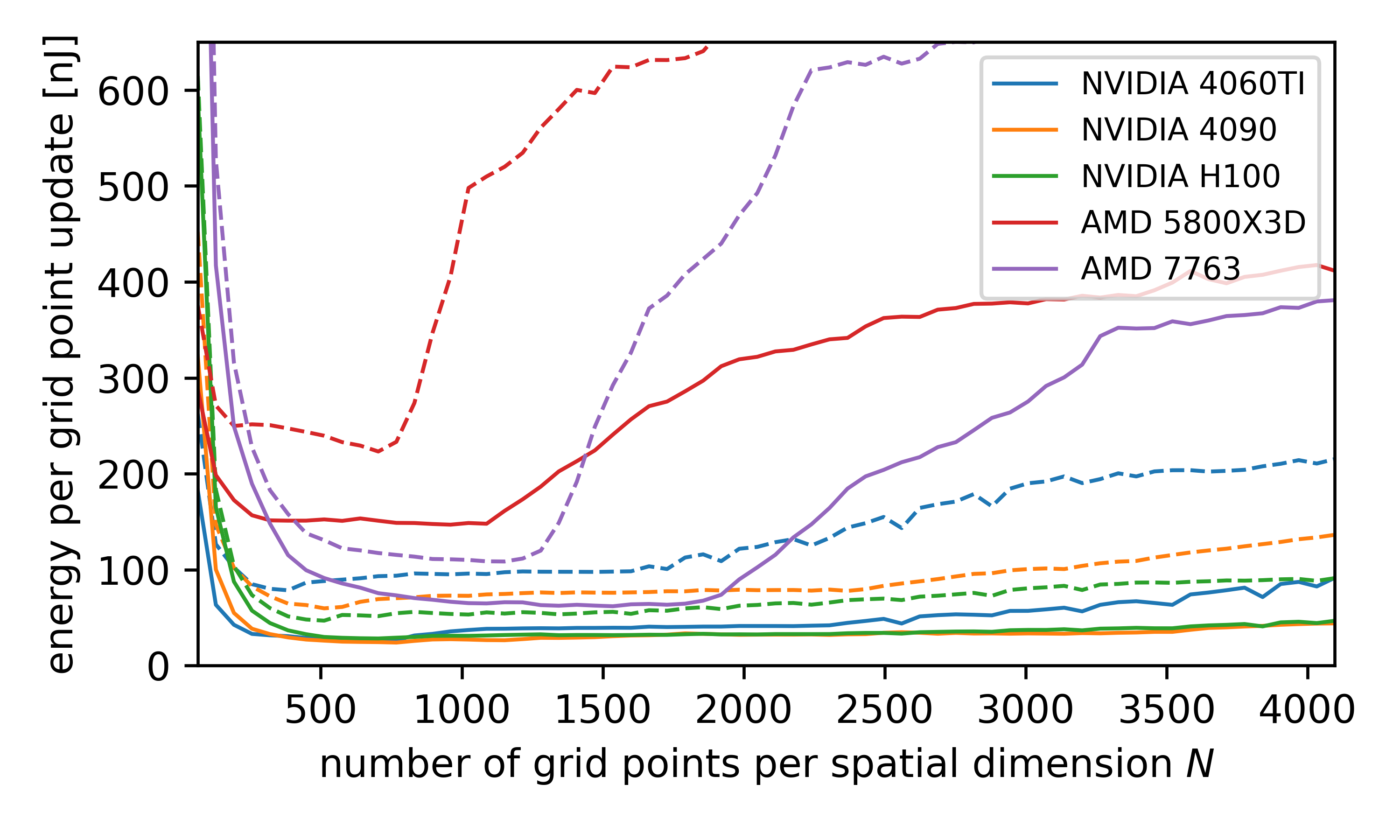}
    \caption{Energy usage per grid point update $E_\mathrm{point}$ for the devices listed in Tab.~\ref{tab:hardware} for different grid sizes for single-precision (solid lines) and corresponding double-precision calculations (dashed lines). For each grid size, the best energy efficiency of different subgrid settings is shown.}
    \label{fig:perf_energy}
\end{figure}

In addition to time-to-solution also the energy-to-solution becomes more and more relevant in high-performance computing applications and computational research. A useful metric is the energy usage per grid point update $E_\mathrm{point}$. It determines the energy $E_\mathrm{sim}$ used for a full simulation on a $N_x \times N_y$ grid with $N_\mathrm{iter}$ time steps as
\begin{align}
    E_\mathrm{sim}&=E_\mathrm{point} N_xN_yN_\mathrm{iter}.
\end{align}
Figure~\ref{fig:perf_energy} shows the measured energy usages per grid point update for the different devices in single and double precision simulations using RK4. For CPU runs, the power consumption is measured as the CPU-socket power consumption with RAPL counters (Running Average Power Limit). For GPU runs only the power consumption of the GPU itself according to the NVIDIA system management interface is used because the performance impact of the CPU in GPU-based simulations with PHOENIX is negligible.

For small grid sizes, i.e., latency-bound situations, the power consumption during the simulation is dominated by the idle power consumption of the devices. Consequently, the NVIDIA 4060 TI can outperform other GPUs in this region due to its low idle power consumption. GPUs show only a weak dependency on the grid size for large grid sizes compared to CPUs. For the large grid sizes shown in Fig.~\ref{fig:perf_energy}, i.e., up to $N_x=N_y=4096$ the NVIDIA 4090 outperforms the NVIDIA H100 in terms of energy efficiency because the subgrid decomposition can efficiently use the large available L2 cache (72 MiB) to avoid the memory bandwidth bound. In the limit of very large grid sizes, the energy efficiency can be estimated from the ratio of the maximal power consumption to the memory bandwidth of the devices.

The lowest energy per grid point update of $24$-$30$ nJ at $N\approx 600$ for single-precision simulations is similar among the tested GPUs. The NVIDIA 4090 shows the highest single-precision energy efficiency of $24$ nJ per grid point update. In double precision, the NVIDIA H100 (46 nJ) is more energy efficient than the NVIDIA 4060 TI (79 nJ) and the NVIDIA 4090 (58 nJ).

For CPU-based simulations, the behavior of the energy per grid point update has three regions similar to the grid point update rates: For small grid sizes the simulation is latency-bound and the power consumption mostly stems from the idle power consumption.  Upon increasing the grid size the idle power gets less dominant compared to the dynamical power and a plateau of power consumption is entered where the simulation is bound by cache transfers instead of latency. In this intermediate grid size, e.g., $400 \leq N \leq 1800$ for the AMD 7763 CPU the CPUs can achieve their optimal power efficiency with PHOENIX of about $60$ nJ per grid point update for single-precision simulations. Beyond the cache limit, the power consumption increases rapidly due to the memory bandwidth bound.

\subsection{Conclusion}\label{sec:perf_conclusion}
The performance model and results detailed in this section lead to the conclusion that entry-to-midrange consumer GPUs can be recommended for simulations with PHOENIX. Device architectures with large L2 caches are especially advantageous for simulations. However, data-center GPUs like the NVIDIA H100 are faster and more efficient for double-precision simulations. It should also be noted that due to their integration into HPC environments, their overall efficiency including cooling and maintenance expenses can be significantly better than for a large setup consisting of air-cooled consumer GPUs.

\section{Applications}\label{sec:application}
In the following, several applications of PHOENIX are presented. These examples are included with the code \cite{phoenix_release} and can be run using the corresponding MATLAB and Python scripts. The selection of examples is designed to illustrate the capabilities of PHOENIX and to cover various variations of the equations implemented as well as a number of different variants, types, and possibilities of post-processing that can be performed. Applications include extensions of our previous research \cite{Ma2016, PhysRevB.101.205301, PRXQuantum.2.030320, Li2022, schneider2024topological} covering the fields of topology, non-Hermitian physics, nonlinear physics, and quantum-state tomography based on Monte Carlo simulations with quantum noise. This relatively broad spectrum of examples is intended to help the easy and widespread use of PHOENIX and its adaptation to other interesting scenarios. These examples also showcase the exceptional performance and adaptability of the PHOENIX code and its potential on high-end consumer hardware for a wide range of practical applications.

First, the full set of equations covered by PHOENIX is introduced. For widespread applicability, we include a number of extensions going beyond Eq. (\ref{eq:1}) that was used for benchmarking purposes above. This enables the application of PHOENIX to a wider class of effectively two-dimensional systems with coherent driving or incoherent source or excitation as well as to other nonlinear systems or condensate systems described by a variation or subset of Eq. (\ref{eq:PULSE_psi}) and (\ref{eq:PULSE_res}). Including the extensions to the dNLSE in Eq. (\ref{eq:1}) currently implemented, the set of equations covered by PHOENIX reads:

%\iftwocolumn
%\begin{widetext}
%\begin{align}
%    d\psi_\pm= \frac{-i}{\hbar}\left(\biggl(-\frac{\hbar^2}{2m}\nabla^2+V+g|\psi_\pm|^2+g_\mathrm{x}|\psi_{\mp}|^2+g_\mathrm{r}n_\pm + \frac{i\hbar}{2}[Rn_\pm-\gamma] \biggr) \psi_\pm+J_\pm\psi_{\mp}+E_\pm\right)dt+dW,
%\label{eq:PULSE_psi}
%\end{align}
%\begin{align}
%    \frac{d n_\pm}{d t} = P_{\pm} - (\gamma_\mathrm{r}+R|\psi_\pm|^2)n_\pm.
    %\partial_\mathrm{t}n_\pm = P_{\pm} - (\gamma_\mathrm{r}+R|\psi_\pm|^2)n_\pm.
%\label{eq:PULSE_res}
%\end{align}
%\end{widetext}
%\else
\begin{align}
    &d\psi_\pm=\nonumber \\
    &-\frac{i}{\hbar}\left(-\frac{\hbar^2\nabla^2}{2m}+V+g|\psi_\pm|^2+g_\mathrm{x}|\psi_{\mp}|^2+g_\mathrm{r}n_\pm + \frac{i\hbar}{2}[Rn_\pm-\gamma] \right) \psi_\pm dt  \nonumber \\
    &-\frac{i}{\hbar}\left(J_\pm\psi_{\mp}+E_\pm\right)dt+dW,
\label{eq:PULSE_psi}
\end{align}
\begin{align}
    \frac{d n_\pm}{d t} = P_{\pm} - (\gamma_\mathrm{r}+R|\psi_\pm|^2)n_\pm.
    %\partial_\mathrm{t}n_\pm = P_{\pm} - (\gamma_\mathrm{r}+R|\psi_\pm|^2)n_\pm.
\label{eq:PULSE_res}
\end{align}
%\fi

Here $\psi_\pm=\psi_\pm(\textbf{r},t)$ is a complex-valued spinor field, with $\pm$ denoting the two spin or polarization components and $n_\pm=n_\pm(\textbf{r},t)$ are spin-dependent densities of incoherent excitation reservoirs. Parameters $g$ and $g_\mathrm{x}$ scale the self- and cross-interaction of the two spin-components of $\psi_\pm$, respectively, while $g_\mathrm{r}$ scales the interaction with excitation reservoirs $n_\pm$ (here included in equal spin channels, which can be generalized). The gain to $\psi_\pm$ from the reservoir scales with $R$ while driving by resonant coherent excitation is included through $E_\pm=E_\pm(\textbf{r},t)$. If not independent from each other, the two spin components are coupled for example through $J_\pm= \Delta_\mathrm{LT}(\partial_x \mp i\partial_y)^2$, here included to introduce a typical spin-orbit interaction from transverse-longitudinal splitting of the cavity field with magnitude $\Delta_\mathrm{LT}$. $P_\pm=P_\pm(\textbf{r})$ is an external pump that replenishes the reservoir, while $\gamma_\mathrm{r}$ introduces loss from the reservoir. The complex Wiener noise is denoted by $dW$. When PHOENIX is executed, the boundary conditions in the $x$- and $y$-directions can be set independently peridoic or zero. This allows to apply the truncated Wigner approximation and to investigate classical and quantum fluctuations \cite{lindberg1988effective, takayama2002t, wouters2007excitations, wingenbach2022dynamics, PRXQuantum.2.030320}.

We note that Eqs.~(\ref{eq:PULSE_psi}) and (\ref{eq:PULSE_res}) can describe a variety of two-dimensional physical systems governed by variants of the NLSE. Besides the polariton system discussed below this also includes systems such as conventional Bose-Einstein condensates (BECs), superfluids, and quantum gases. 
The introduction of additional terms going beyond Eq. (\ref{eq:PULSE_psi}) and (\ref{eq:PULSE_res}) is user-friendly and detailed instructions are available in the PHOENIX documentation. In the following, multiple examples for potential applications that also require different types of pre- and post-processing are illustrated and the code to run these examples, including easily accessible Python and Matlab scripts, respectively, that call the pre-compiled main numerical PHOENIX routines, is provided together with the PHOENIX release \cite{phoenix_release}.

As a prominent application area of the set of Eq.~(\ref{eq:PULSE_psi}) and (\ref{eq:PULSE_res}), in the following we will focus on the description of exciton-polaritons in planar microresonators. Exciton-polaritons are hybrid light-matter quasi-particles resulting from the strong coupling of exciton and cavity photons in semiconductor microcavities~\cite{kavokin2017microcavities}. For both resonant and nonresonant optical excitation of the cavity, macroscopic coherence of polaritons can form known as polariton condensation~\cite{deng2002condensation, Kasprzak2006, RevModPhys.82.1489}. In case of nonresonant excitation, the interaction of the condensate and the excitation reservoir can be used to induce a repulsive potential energy landscape for the condensate polaritons to move in, allowing for optical trapping and control~\cite{wertz2010spontaneous, sanvitto2011all, PhysRevLett.109.216404, PhysRevLett.110.186403, PhysRevB.88.041308}. In materials with larger exciton binding energies polariton condensation even occurs at relatively high temperatures up to room temperature~\cite{christopoulos2007room, Kedziora2024}. In contrast to conventional BECs, these condensates are inherently non-Hermitian as a constant supply of particles is required to compensate for the finite excitonic and polaritonic lifetimes. The nonlinear nature of polariton condensates leads to a variety of intriguing phenomena such as the formation of solitons and vortices~\cite{byrnes2014exciton}. Furthermore, multistabilities in polariton condensates allow for novel designs of all-optical transistors~\cite{baas2004optical, gippius2007polarization, paraiso2010multistability, ballarini2013all}. The strong nonlinearity in polariton condensates enables new designs for neuromorphic networks~\cite{Mirek2021}. Due to the two active exciton spin states coupled to the two circular polarizations of the cavity light field, polaritons possess a spin degree of freedom, which permits the emergence of new vortex states, such as half- and spin-vortices~\cite{PhysRevLett.99.106401,lagoudakis2009observation,PhysRevB.101.205301}. The splitting of circular polarization modes into TE and TM modes that typically occurs in planar photonic resonators leads to spin-orbit coupling within the condensate~\cite{PhysRevB.59.5082}. This permits the observation of interesting phenomena such as the optical spin Hall effect~\cite{PhysRevLett.95.136601,leyder2007observation,Lafont2017,Luk2021} and oblique half-dark-solitons~\cite{PhysRevB.83.193305,hivet2012half}. Polaritons in the different spin components also interact with each other, inducing a nonlinearity that can be attractive or focussing in nature \cite{PhysRevB.82.075301}. Under resonant excitation, squeezed states were realized~\cite{Boulier2014} and quantum correlations of the light source were mapped to the matter system in polariton condensates~\cite{PhysRevLett.115.196402}. In Refs.~\cite{PRXQuantum.2.030320, luders2023tracking, luders2023continuous} the quantum coherence of a non-resonantly excited polariton condensate was quantified, illustrating polariton condensates as a promising hybrid light and matter platform also for quantum information processing.

\subsection{Example 1: Coherent excitation of polariton topological corner states in large AAH-SSH potential lattices}
\label{cha:ex1}
Over the past years, numerous studies have focused on the formation and utility of topological states in both electronic condensed matter and optical systems, for the latter including photonic crystals, coupled waveguides, meta-surfaces, and exciton-polaritons. In two dimensions topologically protected states may manifest as edge or corner states, respectively. A well-known example of a structure in which these states appear is the Su-Schrieffer-Heeger (SSH) model in 1D and 2D~\cite{st2017lasing, benalcazar2017quantized}. The topologically protected states are highly robust~\cite{PhysRevB.100.075437} which makes them promising candidates for example for unidirectional signal propagation~\cite{wang2009observation}, optical communication, stable single-mode lasing, and quantum information processing~\cite{st2017lasing}. Another structure recently investigated are coupled AAH-SSH potential lattices in which both edge and corner states may be observed \cite{schneider2024topological}. In two dimensions, structures investigated typically consist of a domain in space with a large number of potential wells \cite{Klembt2018}, rendering full numerical investigations of large structures time-consuming. 

\begin{figure}
    \centering
    \includegraphics[width=1.0\columnwidth]{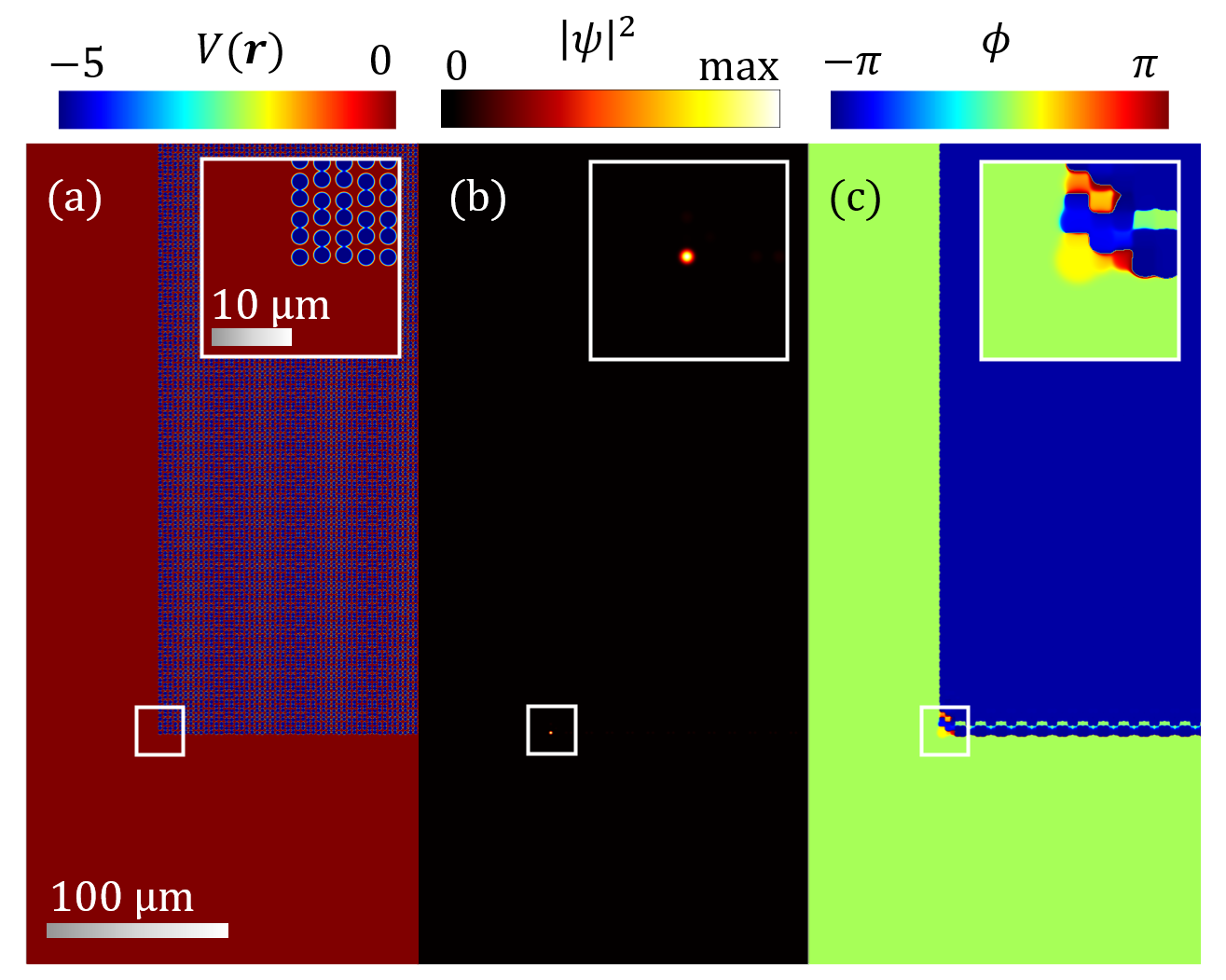}
    \caption{(a) Extended double-wave lattice with $9000\times9000$ grid points and $272\times272$ potential wells. (b-c) Density and phase of topological corner state localized in the corners of the real space grid. For illustrative purposes, a segment around the bottom left corner of the potential lattice is displayed. The insets show a zoom onto one corner of the potential lattice and the corresponding corner state.}
    \label{fig:example1}
\end{figure}

In Fig.~\ref{fig:example1} we show results for an example in which a significantly larger structure than in Ref.~\cite{schneider2024topological} is investigated, enabled by the efficiency of PHOENIX. We demonstrate that the corner states discussed in Ref.~\cite{schneider2024topological} are indeed also stable in much larger systems of the same AAH-SSH structure. The physical parameters used to solve Eq.~(\ref{eq:PULSE_psi}) are given in Ref.~\cite{parameters_example1}. Parameters set to zero reduce the set of equations, Eq.(\ref{eq:PULSE_psi}) and (\ref{eq:PULSE_res}), to the subset of equations required to describe the physical setup in the present example. Given a very large grid size of $N\times N = 9000 \times 9000$ grid points, the resulting system state vector propagating in time covers $0.81\times10^8$ distinct elements. Due to the large GPU memory-band width, however, calculations on systems of this size are now very much feasible. For the version of Eq.~(\ref{eq:PULSE_psi}) used in this example we note that the maximum grid size compatible with the total memory of 24GB of one single RTX 4090 GPU is about $~2.25\times10^8$ (corresponding to a $15000\times 15000$ square grid).

The polariton condensate is excited coherently to drive the system near the frequency of a pre-determined corner state. This frequency is determined by eigenvalue calculations for the linearized system operator, which is numerically quite costly and does not deliver the correct resonance frequency in the nonlinear regime. Utilizing the PHOENIX speedup in the time-domain calculations, optimizing the driving frequency is feasible using local optimizers like \textit{fmindbnd}. The optimized coherent pump underlying Fig.~\ref{fig:example1} features a frequency of $\omega = -5.6883~\mathrm{THz}$ and an amplitude of $E_0 = 5.6856~\mathrm{\upmu eV}$. Analytically, the potential well distribution in the 2D double-wave structure is given by
\begin{equation}
    V(x,y) = \sum_n V_0\mathrm{exp}\left(-\frac{(x-x_n)^2+(y\pm y_n)^2}{w_v^2}\right)^{10}
    \label{eq:example1_pot}
\end{equation}
as depicted in Fig.~\ref{fig:example1}(a). Each of the $272$ potential wells in $x-$ and $y-$direction feature a depth of $V_0 = -5~\mathrm{meV}$ and a width of $w_v = 1~\mathrm{\upmu m}$.

After $5~\mathrm{ns}$ the system converges into the corner state displayed in Fig. \ref{fig:example1}(b) and (c). Note that for illustrative purposes only a segment around the bottom left corner of the potential lattice is shown. Using AMD EPYC 7443P CPU and execution on the NVIDIA RTX 4090 GPU, time evolution of the corner state for a given frequency takes about 7.5 hours in real time. Here, the subgrid decomposition is used to minimize the overhead. A total number of $18\times 18$ subgrids are used by setting the flag \textit{--subgrids 18 18} when executing PHOENIX.

\subsection{Example 2: Algorithm-assisted localization of exceptional points in reservoir-coupled polariton condensates}
\label{cha:ex2}
In recent years, there has been a growing interest in investigating non-Hermitian phenomena in dissipative systems. In these open systems, loss and gain lead to complex eigenvalues and potentially nonorthogonal eigenvectors of the operator governing the time evolution of a given system~\cite{kato2013perturbation, wiersig2019nonorthogonality}, permitting the observation of striking phenomena like the non-Hermitian skin effect or exceptional points~\cite{berry2004physics, bender2007making}. The latter are singularities at which two or more eigenvalues are degenerate and their respective eigenvectors become parallel~\cite{heiss2004exceptional, heiss2012physics}. Exceptional points have been investigated in a variety of physical systems like microwave resonators~\cite{dembowski2001experimental}, atomic systems~\cite{choi2010quasieigenstate}, plasmonic nanostructures~\cite{kodigala2016exceptional}, optical waveguides~\cite{guo2009observation}, microresonators~\cite{lee2009observation}, and non-reciprocal systems~\cite{fruchart2021non}. Near the exceptional point, the coalescence of the eigenvectors leads to counterintuitive system behavior, such as loss-induced transparency~\cite{zhang2018loss}, unidirectional invisibility and reflectivity exceeding unity~\cite{lin2011unidirectional}, and lasing self-termination~\cite{el2014exceptional}. The exceptional point order $\eta$ is determined by the number of eigenvectors and eigenvalues coalescing at the exceptional point~\cite{chen2017exceptional}. Exceptional points are promising candidates for a new generation of sensors, as the frequency splitting near exceptional points scales with the $\eta^\mathrm{th}$-root of the perturbation and thus outperforms conventional Hermitian sensors~\cite{wiersig2020review, mandal2021symmetry}.

Due to their non-equilibrium nature, polariton condensates offer a natural playground to study non-Hermitian phenomena such as exceptional points. By brining two modes into resonance with each other and fulfilling the condition that the mode coupling corresponds to the modes' gain difference, exceptional points have been realized in polariton condensates in the past~\cite{gao2015observation, gao2018chiral, hanai2019non, Li2022, PhysRevB.109.085311}. Moreover, the nonlinearity of polariton condensates allows for the investigation of the interplay of nonlinear and non-Hermitian physic~\cite{wingenbach2024manipulating}. However, localizing $\eta^\mathrm{th}$ order exceptional points in parameter space can be challenging, as $2\eta-2$ parameters require precise adjustment~\cite{mandal2021symmetry}. In nonlinear systems, this becomes even more challenging as the nonlinear characteristics such as the saturable gain and Kerr-nonlinearities couple mode gain, loss, and energy~\cite{wingenbach2024manipulating}. In the following, we show an example where the short computation times of PHOENIX permit the application of numerical optimization algorithms to localize exceptional points in parameter space consuming only a relatively small amount of time. 

In Ref.~\cite{Li2022}, we realized an optical switch for the polariton condensation in a double-well potential featuring a second-order exceptional point. The double-well potential is given by
\begin{equation}
    V(x,y)=V_1 f(x_1,w_v)+V_2 f(x_2,w_v)
    \label{eq:example2_pot}
\end{equation}
with the envelope function $f(x_\mathrm{shift},w)= \mathrm{exp}\left(-\sfrac{\left[(x-x_\mathrm{shift})^2+y^2\right]}{w^2}\right)^2$
Here, $w_v=1.5~\mathrm{\upmu m}$ describes the radius of the potential wells. Furthermore, $V_1=-2.2$ and $V_2=-2~\mathrm{meV}$ define the depths of the left and right potential well, respectively. Their distance is described by $d=x_2-x_1=4$ $\mathrm{\upmu m}$, which directly influences the mode coupling of the condensates confined by the individual potential wells. The system is driven by nonresonant excitation with one Gaussian-shaped pump in each potential well:

\begin{equation}
    P(x,y)=P_1 f(x_1,w_p)+P_2 f(x_2,w_p).
    \label{eq:example2_pump}
\end{equation}

\begin{figure}
    \centering
    \includegraphics[width=1.0\columnwidth]{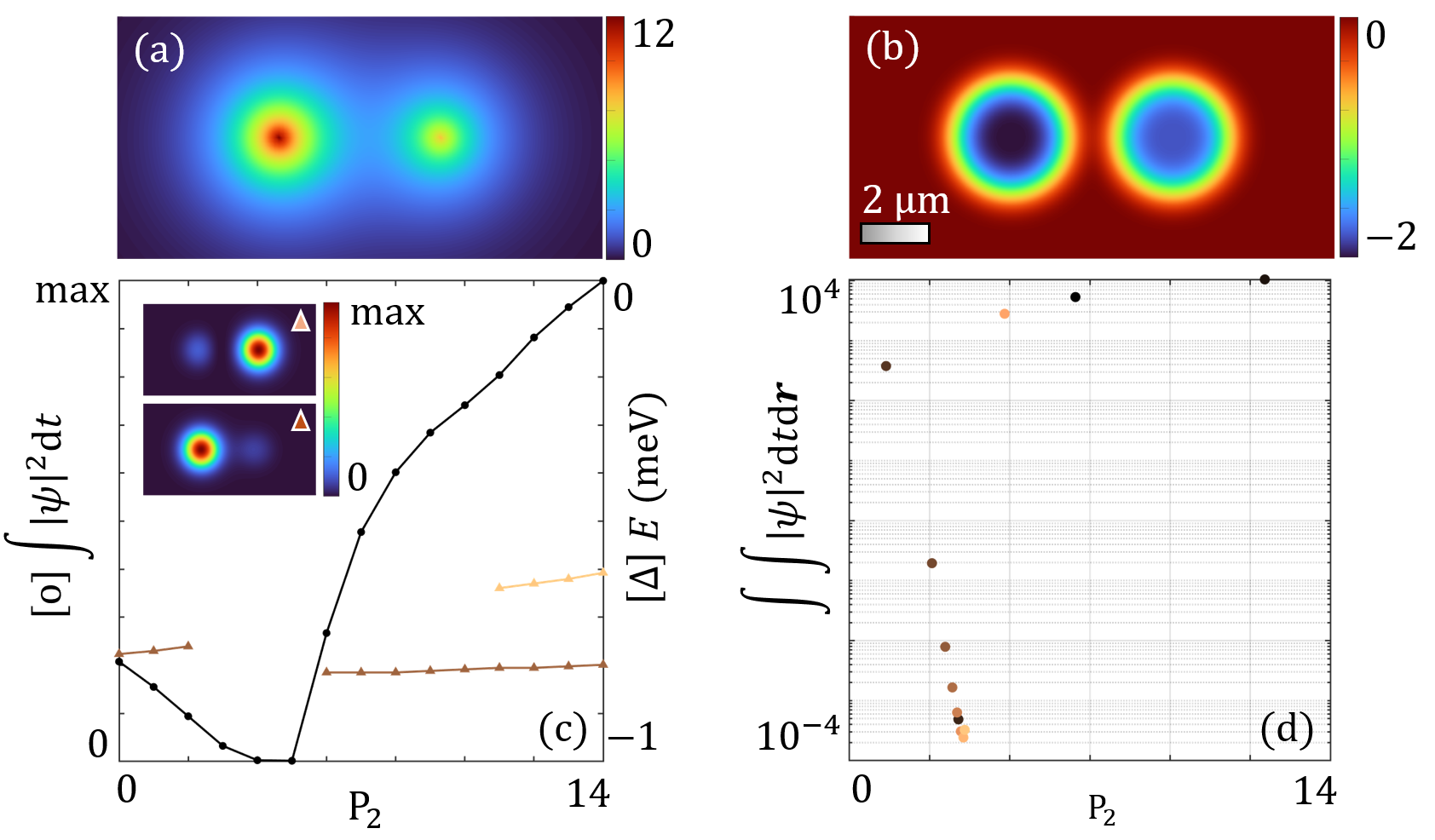}
    \caption{(a) Spatial pump in $\mathrm{ps^{-1}\upmu m^{-2}}$ and (b) double-well potential distribution in $\mathrm{meV}$. (c) The integrated density of the polariton modes in $\mathrm{\upmu m^{-2}}$ (inserted plots) in black and the mode energies in scales of brown, indicating the condensation switch-off and the energy bifurcation at the exceptional point. Note that the second mode passes the condensation threshold at $P_2>10~\mathrm{ps^{-1}\upmu m^{-2}}$. (d) Algorithm-assisted localization of the exceptional point using local optimizer "fmindbnd" on an intensity interval $P_2=[0; 20]~\mathrm{ps^{-1}\upmu m^{-2}}$. The increasing iteration number is illustrated by the marker color (from dark to light).}
    \label{fig:example2}
\end{figure}

$P_1=12~\mathrm{ps^{-1}\upmu m^{-2}}$ defines the pump intensity in the left, while $P_2$ defines the intensity in the right well. The pump width is given by $w_p=1~\mathrm{\upmu m}$ and their separation is set to $d=x_2-x_1=4~\mathrm{\upmu m}$. The choice of parameters used to solve Eq. (\ref{eq:PULSE_psi}) and (\ref{eq:PULSE_res}) are provided in Ref.~\cite{parameters_example2}. Parameters set to zero reduce the set of equations, Eq.(\ref{eq:PULSE_psi}) and (\ref{eq:PULSE_res}), to the subset of equations required to describe the physical setup in the present example. The spatial distribution of pump and potential are shown in Fig.~\ref{fig:example2}(a-b). The two lowest-energy eigenmodes are taken as the initial condition for the system evolution. These are the binding and anti-binding modes of the potential wells, which are focused in the left or right well respectively [see insets in Fig.~\ref{fig:example2}(c)]. The pump intensity $P_2$ is then gradually increased, which not only shifts the gain and loss difference between the two modes, but also the energy difference due to a nonlinear blueshift. In Fig.~\ref{fig:example2}(c) the tracing of the exceptional point in dependence of the pump intensity $P_2$ is illustrated.
Here, increasing the pump intensity initially leads to a counterintuitively decreased integrated density in the system and the energy spectrum shows only the bonding mode. At the exceptional point, the density falls below the condensation threshold leading to a rapid increase in the mode linewidth and thus no observable energy spectrum in the numerics. Further increasing the pump intensity leads to recondensation and appearance of the antibonding mode in the energy spectrum above its threshold of $P_\mathrm{thr} = 10~\mathrm{ps^{-1}\upmu m^{-2}}$. The mode energy bifurcation and condensate switching-off is in good agreement with the experimental data shown in Ref.~\cite{Li2022}. Due to the PHOENIX speedup, it is now feasible to localize the exceptional point automatically with optimizer algorithms. In  Fig.~\ref{fig:example2}(d) the exceptional point is localized with the MATLAB built-in one-dimensional minimizer for a pump intensity interval $P_2=[0; 20]~\mathrm{ps^{-1}\upmu m^{-2}}$. Here, the minimizer finds the local minimum in the integrated density to localize the EP. The iteration number is illustrated by the lightening of the marker color (dark to light, marks an increase in the iteration number). The very fast evaluation of each step in the optimization process promises feasibility also for other tasks requiring optimization or complex inverse design \cite{Rockstuhl_inverse_polaritons} in a complex multi-dimensional parameter space and could be adapted to wavefunction tuning by adjusting the system parameters to minimize the discrepancy between the resulting density and the desired density profile. Using an AMD EPYC 7443P CPU and NVIDIA RTX 4090 GPU the EP localization takes about 1 hour in real time.

\subsection{Example 3: Localized vortices in spinor polariton condensates with spin-orbit coupling}
\label{cha:ex3}
In this chapter, we investigate the multistability of localized vortices in spinor polariton condensates, as done in Ref.~\cite{PhysRevB.101.205301}. Quantized vortices are a well-studied phenomenon in conventional Bose-Einstein condensates and driven-dissipative polariton condensates~\cite{lagoudakis2008quantized, sanvitto2010persistent, roumpos2011single,Liew2015}. Quantized vortices are characterized by a funnel-shaped deepening and a phase singularity in the so-called vortex core surrounded by a phase winding of $2\pi m$. Here, the integer number $m$ is referred to as the topological charge of the vortex. In the case of polariton condensates the vortex results in a quantized orbital angular momentum (OAM) which can be extracted from the cavity-emitted light by interferometry methods~\cite{baranova1983wave}. Thanks to that, quantized vortices in polariton condensates are promising candidates for novel information processing and data storage applications, in which their topological charge act as bit values~\cite{ma2017vortex} with the potential also for qubit analogue realizations \cite{Xue2021,Barrat2024}. Hence, recent research on polariton condensates aims at reliable trapping and manipulating of quantized vortices. For example, quantized vortices can be trapped, moved, and switched between different topological charges by designing optically-induced potentials arising from the repulsive interaction with the excitonic reservoir ~\cite{PhysRevLett.106.115301, ostrovskaya2012dissipative, ma2020realization, wingenbach2022dynamics}.

Polariton condensates exhibit a spin degree of freedom, which results from the two optically active exciton spin states which are coupled to the two circular polarization states of the light field. This gives rise to novel vortex states such as half-quantum vortices, i.e. a vortex in one spin component and a non-vortex state in the other ($m_\pm\neq0 \wedge m_\mp=0$)~\cite{seo2015half, seo2016collisional}, as well as full- and spin-vortices, featuring parallel ($m_\pm=m_\mp$) or antiparallel ($m_\pm=-m_\mp$) circulating vortices in the two spin components~\cite{manni2013hyperbolic, wingenbach2022dynamics}. Here $m_\pm$ denotes the topological charge of the respective spin component of the condensate. Vortices in different spin components are coupled via so-called TE-TM (or longitudinal-transverse) splitting, which describes the energy splitting of perpendicularly polarized cavity photon modes and induces a spin-orbit interaction of polaritons.

Investigating spinor polariton condensates demands solution of the dNLSE shown in Eq.~(\ref{eq:PULSE_psi}) and (\ref{eq:PULSE_res}) for both spin components $\pm$ for the polariton field and the reservoir which doubles the dimension of the numerical problem to be solved; in addition smaller time steps may be required to suitably resolve the system dynamics including the spin-orbit interaction operator. Therefore, a resource-efficient implementation such as PHOENIX is needed to allow for a time-efficient and systematic scanning of different vortex states while varying the topological charge configuration. By adding \textit{-tetm} to the runstring, PHOENIX solves the set of coupled equations for the two spinor components $\psi_\pm$ of the polariton condensate and their respective reservoirs (for simplicity here assuming spin preservation in the relaxation from the reservoir).

In the numerical simulations, we use a $x$-linearly polarized continuous-wave pump with a ring-shaped profile
\begin{align}
    P_\pm(\textbf{r}) = P_0\frac{\textbf{r}^2}{w_p^2}\mathrm{exp}\left(\frac{\textbf{r}^2}{w_p^2}\right),
\end{align}
with a radius $w_p$ and a pump intensity $P_0=100~\mathrm{\upmu m^{-2}ps^{-1}}$. The choice of parameters used to solve Eq.~(\ref{eq:PULSE_psi}) and (\ref{eq:PULSE_res}) are provided in Ref.~\cite{parameters_example3}. Parameters set to zero reduce the set of equations, Eq.(\ref{eq:PULSE_psi}) and (\ref{eq:PULSE_res}), to the subset of equations required to describe the physical setup in the present example.

\begin{figure}
    \centering
    \includegraphics[width=1.0\columnwidth]{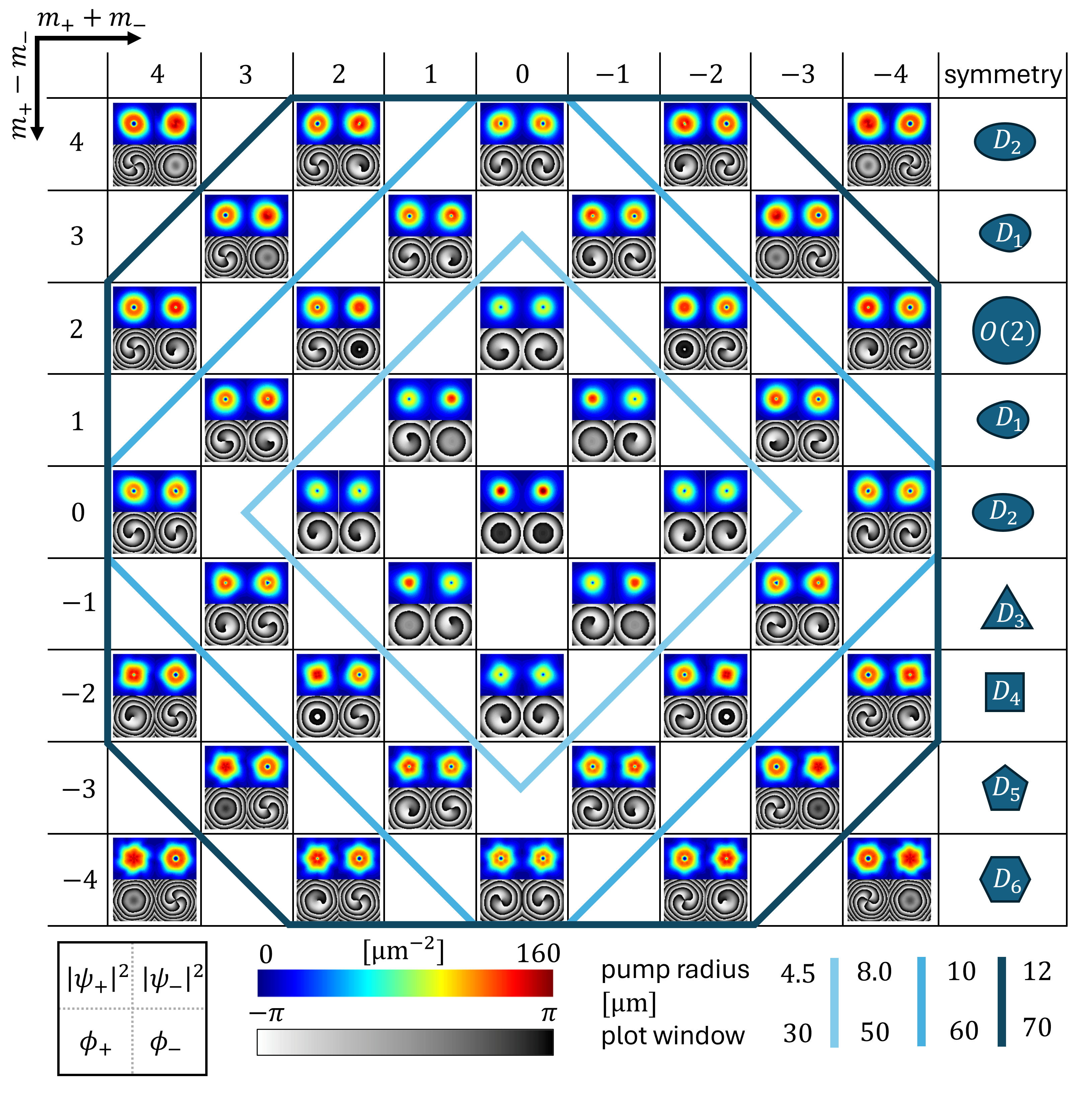}
    \caption{Overview of the first seven groups of approximate real-space symmetry for different stable vortex states. The states are arranged according to the topological charge difference $m_+-m_-$ (vertical) and the sum $m_++m_-$ (horizontal). The resulting symmetries are exemplified on the right-hand side. For each state, the densities and phase distributions of the circular polarizations are shown as in the scheme at the bottom left. The respective color bars are located at the bottom, together with the pump radii and the sizes of the plot windows used for the respective regions marked in blue.}
    \label{fig:example3}
\end{figure}

The dynamically stable vortex states are depicted in Fig.~\ref{fig:example3} together with their topological charge difference $m_+-m_-$ and sum $m_++m_-$, as well as for different pump widths $w_p$. The spin-orbit interaction breaks the symmetry of the vortex states leading to a variety of real-space symmetries in the polariton density. We emphasize that in this application the high computational speed of PHOENIX enables the systematic investigation and classification of a large number of vortex shapes in the spinor system in a remarkably short time. Here, the temporal evolution over $10\,\mathrm{ns}$ is evaluated for each subplot to ensure convergence on a medium-sized grid of $500\times500$, resulting in a runtime of 1 minute in real time on an AMD EPYC 7443P CPU and NVIDIA RTX 4090 GPU for each individual run, with 41 runs in total for the displayed results.

\subsection{Example 4: Phase-space sampling and quantum-state tomography of polariton condensates with Monte Carlo simulations}
\label{cha:ex4}
In this chapter we show how PHOENIX can be used to analyze statistical properties by taking into account classical and quantum fluctuations. We reproduce the central results in Ref.~\cite{PRXQuantum.2.030320} and determine the quantum coherence as a measure for the amount of Fock-state superpositions within the polariton condensate using the truncated Wigner approximation (TWA). 

Polariton condensates have been extensively studied as examples of spontaneous buildup of macroscopic coherence under nonresonant excitation. The degree of spatial or temporal coherence is usually characterized by normalized correlation functions. Most prominent are the first- and second-order correlation functions $g^{(1)}$ and $g^{(2)}$ which capture phase and intensity correlations. However, the $g^{(1)}$ function does not contain information of a state being classic or quantum, and $g^{(2)}$ only relates to the diagonal elements of the density operator in Fock space. In contrast, the quantum coherence gives a measure for the off-diagonal contributions of the density operator, i.e. the extent of quantum superpositions of particle-number states~\cite{PhysRevLett.101.067404, PhysRevB.81.033307, 10.1063/1.4936889, PhysRevLett.121.047401}.

Coherence properties can be quantified with Monte Carlo methods. By adding \textit{-dW} $\nu$ to the runstring, PHOENIX solves the stochastic differential Eq. (\ref{eq:PULSE_psi}) and (\ref{eq:PULSE_res}) for a renormalized density $\overline{|\psi|}^2=|\psi|^2-(\Delta V)^{-1}$, with $\Delta V = \frac{L^2}{N^2}$ denoting the unit-cell volume of the two-dimensional grid. The correlations of the complex Wiener noise contribution $dW$ follow
\begin{align}
    \langle dW(\textbf{r})dW(\textbf{r'})\rangle &= 0& \nonumber\\
    \mathrm{and}~~\langle dW(\textbf{r})dW^*(\textbf{r'})\rangle &= (Rn+\gamma_\mathrm{c})\frac{\delta_{\textbf{r,r'}}}{2\Delta V}.& 
\end{align}
The parameter $\nu$ in the runstring scales the Wiener noise intensity. For further details, discussion and for other applications of this theoretical approach, please refer to Refs.~\cite{Carusotto2013,Sinatra2002,Wouters2009,PRXQuantum.2.030320,Alnatah2024}. 

Here, we excite a system with a continuous-wave pump with super-Gaussian profile
\begin{equation}
    P(\textbf{r}) = P_0\mathrm{exp}\left(-\frac{\textbf{r}^4}{w_p^2}\right),
\end{equation}

with a width $w_p = 65~\mathrm{\upmu m}$. The choice of parameters used to solve Eq. (\ref{eq:PULSE_psi}) and (\ref{eq:PULSE_res}) are provided in Ref. \cite{parameters_example4}. Parameters set to zero reduce the set of equations, Eq.(\ref{eq:PULSE_psi}) and (\ref{eq:PULSE_res}), to the subset of equations required to describe the physical setup in the present example. To calculate the average polariton condensate excitation number, the stochastic differential equations are solved $150$ times and the wavefunction is sampled every $5~\mathrm{ps}$ for $1~\mathrm{ns}$%for Each data point in the plot (name small time step)
. To ensure convergence the sampling starts after $1~\mathrm{ns}$. Using the sampled data, the expectation values $\langle \hat{n}_k\rangle$, the mean $\langle \hat{n}_c\rangle$ and variance $\langle \left(\Delta\hat{n}_c\right)^2\rangle$ of the averaged polariton excitation number, the second-order correlation function $g^{(2)}(\tau=0)$, and the amount of quantum coherence are determined for different pump intensities $P_0$. A large number of repetitions of the calculations is necessary to ensure precise results of the stochastic evaluations. Here we profit substantially of the high computation speed of PHOENIX. 

\begin{figure}
    \centering
    \includegraphics[width=1.0\columnwidth]{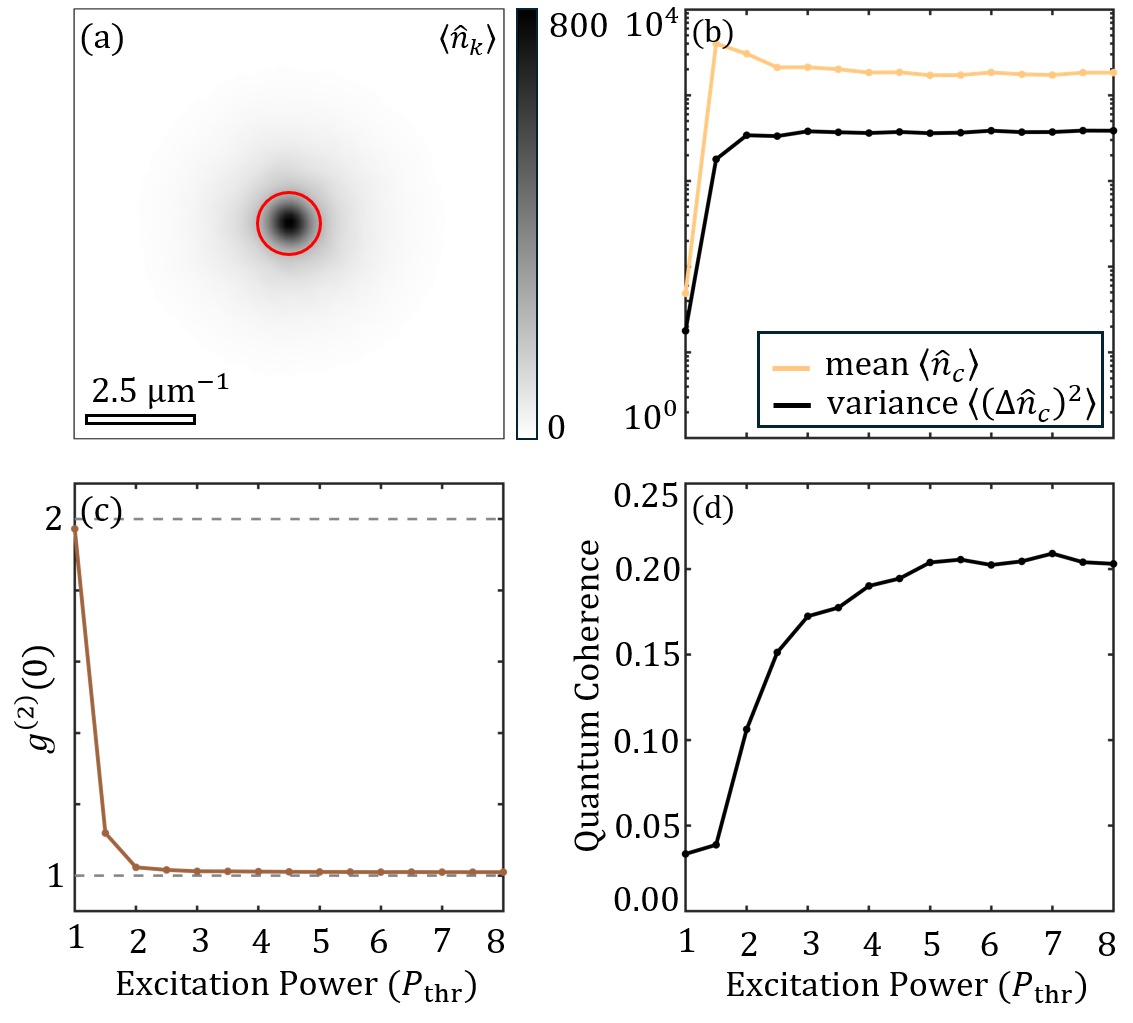}
    \caption{(a) $k$-space expectation value above the condensation threshold $P_0 = P_\mathrm{thr}$. The red square indicates the selected signal for the mode occupation. (b) Mean and variance of averaged polariton excitation number and (c) second-order correlation function $g^{(2)}(0)$ as a function of the excitation power in multiples of the pump threshold. (d) Resulting quantum coherence $\zeta$ in dependence of the excitation power.}
    \label{fig:example4}
\end{figure}

We average the polariton number over a ring covering $N_p$ discrete modes in the vicinity of $k\approx 0$ as displayed by the red ring in Fig.~\ref{fig:example4}(a). The resulting mean polariton excitation number saturates for excitation powers $P_0\geq4P_\mathrm{thr}$; see Fig.~\ref{fig:example4}(b). Fig.~\ref{fig:example4}(c) illustrates the transition of a thermal ($g^{(2)}=2$) to a coherent ($g^{(2)}=1$) state under increasing excitation power. The system's quantum coherence saturates at around $\zeta = 0.2$ as shown in Fig.~\ref{fig:example4}(d). The displayed results are in good agreement with our results in Ref.~\cite{PRXQuantum.2.030320}.

We emphasize that PHOENIX enables the evaluation of stochastic processes even for large two-dimensional systems and high numbers of repetitions. Using AMD EPYC 7443P CPU and NVIDIA RTX 4090 GPU one sampling for the stochastical calculation is performed in about 95 seconds in real time calculated on the GPU. In total 2250 samples are computed for the displayed results, with a total compute time of approximately 60 hours. The target quantities shown in Fig.~10 are then evaluated with the post-processing program included in Ref~\cite{phoenix_release}. Other possible applications would include the actual quantum state tomography and density matrix reconstruction in polariton number space as demonstrated in Refs.~\cite{luders2023tracking,luders2023continuous}.

\section{Conclusion}
In this paper, we have presented the open-source code PHOENIX as a solver for two-dimensional Schrödinger equations with extensions. This solver can be executed and is highly performant on both CPUs and GPUs. For parallelization and efficient cache usage, PHOENIX supports subgrid decomposition. 

We have performed a wide range of benchmarks showing that PHOENIX enables very time- and energy-efficient simulations in single and double precision on entry-class to high-end consumer GPUs. Practically, a workstation configuration with an inexpensive CPU and a few mid-level consumer GPUs can be recommended in case no access to an existing HPC cluster with GPUs is available. For double-precision simulations and on medium to large grids, the lower double-precision floating-point throughput of consumer-grade GPUs limits the performance and efficiency. In this case, datacenter-grade GPUs used in HPC clusters show to be the more time- and energy-efficient option. We note that even in this scenario consumer GPUs exceed the CPU performance by a significant margin.

%We have performed a wide range of benchmarks showing that PHOENIX enables very time- and energy-efficient simulations in single (and only slightly less efficient in double) precision on entry-class to high-end consumer GPUs. Practically, a workstation configuration with an inexpensive CPU and a few mid-level consumer GPUs can be recommended in case no access to an existing HPC cluster with GPUs is available. For double-precision simulations, the lower double-precision floating-point throughput of consumer-grade GPUs limits the performance and efficiency to a small extend so that ultimately datacenter-grade GPUs used in HPC clusters are the more time- and energy-efficient option.

The CPU version of PHOENIX also delivers good performance on intermediate grid sizes and, thus, can be a suitable option if GPUs are not available. It should be emphasized here that PHOENIX is easy to extend by additional terms. In direct comparison to a more conventional Runge-Kutta implementation in MATLAB, PHOENIX shows a speedup of up to three orders of magnitude and energy savings of up to 99.8\%. 

PHOENIX is currently tailored for single-CPU and single-GPU based simulations for grid sizes up to the order of $10^4$ in one of the two spatial dimensions (total grid size on the order of $10^8$). Thus, the natural limit for the grid is the main memory in CPU-based simulations and the GPU memory in GPU-based simulations. For very large grid sizes exceeding the available device or node memory, PHOENIX can be straightforwardly generalized to multi-node or multi-GPU usage by distributing the subgrid decomposition. Although PHOENIX currently uses CUDA for the GPU implementation, care was taken to use only ''portable'' CUDA functionality which makes automated porting to GPU architectures from other vendors straightforward.

Furthermore, we provide a series of examples for describing exciton-polariton condensates including codes that illustrate the use of the different terms already available in PHOENIX. Here we show that the high efficiency of PHOENIX makes computing on large grids, local optimization algorithms, as well as large-scale parameter sweeps and Monte Carlo simulations with quantum noise for large statistical ensembles feasible. With this PHOENIX renders large system sizes and certain types of (such as statistical or iterative) evaluations possible that were previously virtually inaccessible to the typical user. We would further like to point out that in addition to the highly optimized functions described, PHOENIX also masters conventional forms of the imaginary-time algorithm or the split-step Fourier method. For further information on more specific capabilities, please refer to the most recent version of the PHOENIX repository~\cite{phoenix_release}. 

As a natural next step, further equations such as the coupled set of equations of motion to separately handle two-dimensional light field and induced polarization to describe in particular coherent excitations beyond the lower-polariton branch approximation, could be implemented. The presented set of equation can be further extended for example by addtional terms such as the Rashba-Dresselhaus spin-orbit coupling. Moreover, adaptive Runge-Kutta methods or the imaginary-time algorithm could be optimized within the framework of PHOENIX. Since very large grid sizes can be handled by PHOENIX, medium sized three-dimensional real-space grids (on the order of 500 grid points per spatial dimension) could also be investigated with PHOENIX with minor adaptation of the code.

%Author contributions
%https://www.elsevier.com/researcher/author/policies-and-guidelines/credit-author-statement
%David Bauch: Software, Validation, Writing - Original Draft, Writing - Review & Editing,
%Jan Wingenbach: Investigation, Writing - Original Draft, Writing - Review & Editing, Visualization
%Xuekai Ma: Writing - Review & Editing, Resources
%Robert Schade: Software, Validation, Formal analysis, Investigation, Data Curation, Writing - Original Draft, Writing - Review & Editing, Visualization
%Christian Plessl: Supervision, Resources, Writing - Review & Editing
%Stefan Schumacher: Supervision, Conceptualization, Resources, Writing - Review & Editing

%\begin{acknowledgments}
\section*{Acknowledgments}
This work was supported by the Deutsche Forschungsgemeinschaft (German Research Foundation) through the transregional collaborative research center TRR142/3-2022 (231447078, project A04). 

The authors gratefully acknowledge the computing time provided to them on the high-performance computer Noctua 2~\cite{noctua2} at the NHR Center PC2. This is funded by the Federal Ministry of Education and Research and the state governments participating on the basis of the resolutions of the GWK for the national high-performance computing at universities (\url{www.nhr-verein.de/unsere-partner}). The computations for this research were performed using computing resources under project hpc-prf-hdpadi.

We gratefully acknowledge help with implementing the examples in Sec.~\ref{sec:application}; in Sec.~\ref{cha:ex1} from Tobias Schneider, and in Sec.~\ref{cha:ex4} from Franziska Barkhausen and Maximilian Nürmberger. We
thank Hendrik Rose for testing and validating the installation process and core functionalities of PHOENIX.
%\end{acknowledgments}

\bibliographystyle{elsarticle-num}
%\bibliography{sample_noURL_noDOI}

\providecommand{\noopsort}[1]{}\providecommand{\singleletter}[1]{#1}%

\end{document}